\documentclass[a4paper,epsfig]{JHEP3}
\usepackage{cite}
\usepackage{epsfig}
\usepackage{amsfonts}

\usepackage{amsmath}

\def \beq {\begin{equation}}
\def \eeq {\end{equation}}
\def \bea {\begin{eqnarray}}
\def \eea {\end{eqnarray}}
\def \nn {\nonumber}

 \def\e{{\rm e}} 
\def\dd{{\rm d}}   
  
\def\Z#1{_{\lower2pt\hbox{$\scriptstyle#1$}}}
\def\X#1{_{\lower2pt\hbox{$\scriptscriptstyle#1$}}}

\def\ApJ#1{Astrophys.\ J.\ {\bf#1}}

\preprint{hep-th/0512262}
\title{Towards inflation and dark energy cosmologies from modified
Gauss-Bonnet theory}

\author{Ishwaree P. Neupane \\
\upshape Department of Physics and Astronomy, University of
Canterbury, Private Bag 4800, Christchurch 8004, New Zealand and\\
Central Department of Physics, Tribhuvan University, Kirtipur,
Kathmandu, Nepal\\
\upshape {\ttfamily e-mail: ishwaree.neupane@canterbury.ac.nz} }
\author{Benedict M.N. Carter\\
\upshape Department of Physics and Astronomy, University of
Canterbury, Private Bag 4800, Christchurch 8004, New Zealand \\
\upshape {\ttfamily e-mail: bmc55@student.canterbury.ac.nz}  }

\abstract{ We consider a physically viable cosmological model that
has a field dependent Gauss-Bonnet coupling in its effective
action, in addition to a standard scalar field potential. The
presence of such terms in the four dimensional effective action
gives rise to several novel effects, such as a four dimensional
flat Friedmann-Robertson-Walker universe undergoing a cosmic
inflation at the early epoch, as well as a cosmic acceleration at
late times. The model predicts, during inflation, spectra of both
density perturbations and gravitational waves that may fall well
within the experimental bounds. Furthermore, this model provides a
mechanism for reheating of the early universe, which is similar to
a model with some friction terms added to the equation of motion
of the scalar field, which can imitate energy transfer from the
scalar field to matter.\\

 }

\begin{document}


\noindent
\section{Introduction}
Although Einstein's theory has been proven to be remarkably simple
and successful as a classical theory of gravitational
interactions, there are several observational facts which it has
failed to elucidate. These cosmological conundrums include both
cosmic inflation, or a period of accelerated expansion in the
early universe, and a recent acceleration in the expansion of the
universe.

Inflation in the early universe is a very attractive proposal for
explaining the present large scale homogeneity and high degree of
isotropy of the universe (one part in $100,000$), in addition to
the observed spectrum of density perturbations, which is usually
attributed to a scalar field rolling down a shallow potential.
Similarly, the current acceleration of the universe, as indicated
by recent cosmological results~\cite{Bennett03a}, is usually
attributed to some form of cosmic fluid having a large and
smoothly distributed negative pressure, usually called {\it dark
energy} or {\it dark pressure}.

Cosmologists have long wondered why/how the universe has been
recently accelerating: is it due to a pure cosmological constant
term, or due to some sort of negative pressure generated by one or
more dynamical scalar fields, or something else? In recent years,
different explanations have been provided for both inflation and
the current epoch of acceleration: some examples of recent
interest include brane-world modification of Einstein's general
relativity (GR), including a 5d DGP (Dvali-Gabadadze-Porrati)
model~\cite{Dvali:2000a}. The names of the dark energy candidates
run the gamut from $f(R)$ gravity~\cite{Carroll:2003a} (modifying
in a very radical manner the Einstein's GR itself) to ghost
condensates (the idea which ncludes a more or less disguised
non-locality)~\cite{Arkani:2003a}. Many of these new proposals are
pathological and do not appear more appealing than the two long
envisioned alternative models of dark energy: a cosmological
constant~\cite{Weinberg:1988cp} and a slowly varying
$\Lambda$-term~\cite{Peebles:1988}.

The cosmological constant is a pure dark energy or vacuum energy,
while the variable $\Lambda$-term is some kind of exotic matter or
a slowly varying potential of a scalar field, usually referred to
as {\it quintessence}~\cite{Steinhardt:1998}. This last category
can comprise a Casimir energy or vacuum polarization effect from
additional compact or curved non-compact spatial dimensions, which
only weakly couple to ordinary matter {\it in contrast} to most
tentative quintessence models, including
k-essence~\cite{Picon:2000} or curvature quintessence. In a brane
world scenario~\cite{Ish2001kd}, for instance, the vacuum energy
(or the dark energy) may be viewed as a smooth {\it brane-tension}
if our universe is a 3-brane embedded in higher dimensional
spacetimes.

Indeed, the past decade has witnessed significant progress in the
building of inflationary models as extensions of standard
cosmology, to accommodate the effects of dark energy. However,
most of the inflationary type potentials studied in the literature
are picked up in a very {\it ad hoc} fashion, rather than
constructing such a potential as a valid solution of the field
equations that follows, for instance, from low energy string
effective actions. In this paper we initiate work in this
direction.

\section{Action and equations of motion}

For a given scalar field $\sigma$ with a self-interaction
potential $V(\sigma)$, an effective action as the low energy
approximation of a fundamental theory of gravity and fields is
written as
\begin{equation}
\label{action1}
S= \int d^{4}{x} \left[\sqrt{-g}\left(\frac{1}{2\kappa^2}
R-\frac{1}{2} (\nabla\sigma)^2 - V(\sigma)\right)\right],
\end{equation}
where $\kappa$ is the inverse Planck scale $M_P^{-1}=(8\pi
G_N)^{1/2}$ and $\sigma$ is a classical scalar field whose
stress-energy tensor acts like a time-varying $\Lambda$. When
studying the dynamics of an inflationary universe, the choice of
the field potential $V(\sigma)$ is of particular interest.
However, the origin and exact nature of the field $\sigma$ and the
functional form of $V(\sigma)$ that acts as an extra source of
gravitational repulsion (or dark energy) are not precisely known.
A real motivation for the gravitational action of the form
(\ref{action1}) arises from the following interesting observation.
Inflation with the dynamics of a scalar field with a
self-interaction potential, $V(\sigma)$, provides a negative
pressure to drive the accelerating expansion of the
universe~\cite{Linde:1981mu} as well as a mechanism for generating
the observed density perturbations~\cite{Mukhanov:1990me}.

The effective action (\ref{action1}) is found to be remarkably
simple, but it excludes, at least, one important piece, which is
the coupling between the scalar field $\sigma$ and the Riemann
curvature tensor. This is because the field $\sigma$, if its
vacuum expectation value is to describe the size and shape of the
internal compactification manifold, generically couples with the
curvature squared terms~\cite{Tseytlin,Narain92a}. To this end,
one welcomes the idea that a dark energy and its associated cosmic
acceleration is due to a modification of general relativity such
that the scalar field $\sigma$ couples to gravity via the
curvature squared terms in the Gauss-Bonnet (GB) combination. The
effective action for the system may be taken to be
\begin{equation}
\label{dilatonGB} S= \int d^{4}{x}\left[\sqrt{-g}\left(
\frac{1}{2\kappa^2} R-\frac{\gamma}{2} (\nabla\sigma)^2 -
V(\sigma)+(\lambda-\xi(\sigma))\, {\cal G}  +\cdots
\right)\right],
\end{equation}
where ${\cal G} \equiv  R^2-4 R_{\mu\nu} R^{\mu\nu} +
R_{\mu\nu\rho\sigma} R^{\mu\nu\rho\sigma}$ is the Gauss-Bonnet
invariant and $V(\sigma)$ and $\xi(\sigma)$ are general functions
of $\sigma$. Dots represent other possible contribution to the
gravitational action, such as $\zeta(\sigma)
(\nabla\sigma)^4$~\cite{Tseytlin,Callan85a}, which we drop here
for simplicity. This is justified, since the quartic term
$(\nabla\sigma)^4$ usually decays faster than the GB curvature
invariant. The most desirable feature of GB type curvature
corrections is that only the terms which are the second
derivatives of the metric (or their product) appear in the field
equations: a feature perhaps most important in order to make a
theory of scalar-tensor gravity ghost free.

For some, a weakness of this approach to model dark energy may
reside in the standard argumentation for this particular kind of
modification of Einstein's theory, especially when one remembers
that string/M theory predicts not only the fourth derivative
gravitational term, like a GB invariant, but also higher-order
terms. As is known, the field dependence of the coupling
$\xi(\sigma)$ has its origin in the variation of the background
spacetime, and, in a spatially flat spacetime background, the GB
term is subject to a non-renormalization theorem which implies
that all moduli dependent higher loop contributions (e.g. terms
cubic and higher order in Riemann tensor)
vanish~\cite{Narain92a,Antoniadis93a}.

We anticipate that the Gauss-Bonnet invariant ${\cal G}$ decays
faster than the coupling $\xi(\sigma)$ grows, so that the term
$\xi(\sigma) {\cal G}$ is only subdominant in the effective
action. Moreover, the Gauss-Bonnet term is a topological invariant
in four dimensions if $\xi(\sigma)$ is a constant, but not if
$\xi(\sigma)$ is a dynamical variable.

In an influential paper, Antoniadis et al.~\cite{Antoniadis93a}
demonstrated the existence of cosmological solutions which avoid
the initial singularity and are consistent with the perturbative
treatments of the string effective actions; see, e.g.;
~\cite{Easther:1996a} for other generalizations. The essential
ingredient of their method is a field-dependent Gauss-Bonnet
coupling, $\xi(\sigma)$, where $\sigma$ characterizes the overall
size and the shape of the internal compactification manifold. In
the present paper we go one step further by demonstrating with new
exact non-singular solutions that inflation in the early universe,
as well as a cosmic acceleration at late times, may be explained
by introducing a scalar potential $V(\sigma)$ for the modulus
field $\sigma$, in addition to a field dependent Gauss-Bonnet
coupling, $\xi(\sigma)$. We have reported some of the key results
of our work in ref.~\cite{IshBen05d}. The present paper will
further elaborate the details of the model. For mathematical
simplicity, we henceforth define the function $f(\sigma) \equiv
\lambda-\xi(\sigma)$.

While the field potential, $V(\sigma)$, was absent in some
original string amplitude computations,
e.g.~\cite{Narain92a,Tseytlin}, implying that $V(\sigma)$ is a
phenomenologically motivated field potential, it is quite possible
that, in the presence of additional sources (like branes, fluxes),
the string effective action would incorporate a non-trivial field
potential, $V(\sigma)$, as is the case revealed recently from
string theory cosmic landscape~\cite{Giddings:2001a,Kachru:2003a}.
Unfortunately, we do not have a precise knowledge about the field
potential $V(\sigma)$. In string theory context, any such
potential might take into account some non-perturbative effects,
as arising from the effects of branes/fluxes present in the extra
dimensions. In this sense, one may consider our model as {\it
string-inspired}.

An interesting question that we would like to ask is what new
features would a dynamical Gauss-Bonnet coupling introduce, and
how can it influence cosmological evolution? Recently, it was
suggested in~\cite{Nojiri05b} that the action~(\ref{dilatonGB})
with some specific choices of the potential and the scalar-{GB}
coupling, namely,
\begin{equation}
V(\sigma)=V_0 {\rm e}^{-\sigma(t)/\sigma_0}, \quad f(\sigma)=f_0
{\rm e}^{\sigma(t)/\sigma_0},
\end{equation}
may be used, in four dimensions, to explain the current
acceleration of the universe, including the phantom crossing the
phantom divide ($w=-1$), with effective (cosmological constant or
quintessence) equation of state of our universe; also
see~\cite{follow-up} for other interesting generalizations.

In this paper, instead of choosing particular functional forms for
$V(\sigma)$ and $f(\sigma)$, as above, we present exact
cosmological solutions that respect the symmetry of the field
equations which follow from~(\ref{dilatonGB}). The graviton
equation of motion derived from the action~(\ref{dilatonGB}) may
be expressed in the following form (see, for example,
ref.\cite{Rizos00a})
\begin{eqnarray}
0 &=& \frac{1}{2\kappa^2} \left(R_{\mu\nu}-\frac{1}{2} g_{\mu\nu}
R\right) -\frac{\gamma}{2} \left(\nabla_\mu \sigma \nabla_\nu
\sigma -\frac{1}{2} g_{\mu\nu} (\nabla\sigma)^2 \right)
+\frac{1}{2} f(\sigma) (4 X_{\mu\nu}- g_{\mu\nu} {\cal G})
\nonumber
\\
&{}& +\,\frac{1}{2} g_{\mu\nu} V(\sigma) +2(g_{\mu\nu}\nabla^2
-\nabla_\mu\nabla_\nu) (f(\sigma)R) -4 g_{\mu\nu}
\nabla^\lambda\nabla^\rho (f(\sigma) R_{\lambda\rho})-4 \nabla^2
(f(\sigma) R_{\mu\nu}) \nonumber
\\
&{}& +\,4\nabla^\rho\nabla_\mu(f(\sigma)R_{\nu\rho})
+4\nabla^\rho\nabla_\nu(f(\sigma)R_{\mu\rho})
+4\nabla^{(\rho}\nabla^{\lambda)}(f(\sigma)R_{\mu\rho\nu\lambda}),
\end{eqnarray}
where $X_{\mu\nu}\equiv R R_{\mu\nu}+ R_{\mu\rho\sigma\lambda}
R_\nu\,^{\rho\sigma\lambda}-4 R_{\mu}\,^\rho R_{\nu\rho}$. The
equation of motion for the scalar field $\sigma$ is similarly
given by
\begin{equation}\label{scalar-wave-eq}
0=\gamma \nabla^2\sigma - \frac{d V(\sigma)}{d\sigma} + \frac{d
f(\sigma)}{d\sigma}\, {\cal G} .
\end{equation}
Next, we consider a four-dimensional background spacetime defined
by the standard Friedmann-Robertson-Walker metric:
\begin{equation}
ds^2=-dt^2 + a(t)^2 \sum_{i=1}^3 (dx^i)^2.
\end{equation}
In this background,
\beq
{\cal G}=24 H^2 (\dot{H}+H^2),
\eeq
where
$H \equiv \dot{a}/a$ is the Hubble parameter and
$\dot{a} \equiv \frac{da}{dt}$. The quantity $4X_{\mu\nu}-g_{\mu\nu}
{\cal G}$ vanishes. The $(\mu,\nu)= (t,t)$ and $(x,x)$ components
of the field equations have the following forms
\begin{eqnarray}
0&=&-\frac{3}{\kappa^2} H^2 +\frac{\gamma}{2} \dot{\sigma}^2
+V(\sigma) -24\dot{\sigma}H^3\frac{d f}{d\sigma}, \label{gravi1}
\\
0&=& \frac{1}{\kappa^2} \left(2\dot{H}+3 H^2\right)+8 H^2 \left(
\ddot{\sigma} \frac{d f}{d\sigma}+ \dot{\sigma}^2 \frac{d^2
f}{d\sigma^2}\right) +16 H\dot{\sigma} \frac{d f}{d\sigma}
 (\dot{H}+H^2) +\frac{\gamma}{2} \dot{\sigma}^2
-V(\sigma)\,.\nonumber\\
\label{gravi2}
\end{eqnarray}
The time evolution equation for $\sigma(t)$ (cf
equation~(\ref{scalar-wave-eq})) can be written as
\begin{eqnarray}\label{scalar1}
&& 0=-\gamma\left(\ddot{\sigma}+3H\dot{\sigma}\right)\dot{\sigma}
-\frac{\dd V(\sigma)}{\dd t} + \frac{\dd f(\sigma)}{\dd t}\, {\cal
G}\nonumber \\
&& \Rightarrow \quad
\gamma\left(\dot{\sigma}\ddot{\sigma}+3H\dot{\sigma}^2\right)
+\frac{\dd}{\dd t} \left(V(\sigma)-f(\sigma) {\cal G}\right) +
f(\sigma) \frac{\dd {\cal G}}{\dd t}=0 \nonumber \\
&& \Rightarrow \quad \frac{\dd}{\dd
t}\left(\frac{\gamma}{2}\,\dot{\sigma}^2+\Lambda(\sigma) \right) +
6H \left(\frac{\gamma}{2}\dot{\sigma}^2\right)+\delta=0,
\end{eqnarray}
where we have defined
\begin{equation}
\Lambda(\sigma) \equiv V(\sigma)-f(\sigma)\,{\cal G},\quad \delta
\equiv f(\sigma)\,\frac{\dd {\cal G}}{\dd t}.
\end{equation}
We will call $\Lambda(\sigma)$ an effective potential. Due to the
Bianchi identity, one of the field equations
(\ref{gravi1})-(\ref{scalar1}) is redundant and hence may be
discarded without loss of generality. In the limit $f(\sigma)
H^2\to 0$, the action (\ref{dilatonGB}) reduces
to~(\ref{action1}).

The $\delta$ term in~(\ref{scalar1}) may account for the creation
of particles due to time variation of ${\cal G}$. A friction-like
term like this was first introduced in~\cite{Albrecht82a} on
phenomenological grounds and a question was subsequently raised
in~\cite{Kofman97a} about the physical origin of such term; in our
model, the $\delta$ term is a clear manifestation of a non-trivial
coupling (or back-reaction) between the field $\sigma$ and the
time-varying Gauss-Bonnet curvature invariant. This term
represents a clear advantage of~(\ref{dilatonGB})
over~(\ref{action1}) as a cosmological action.

\section{Scalar field as a perfect fluid}

In a spatially flat background, the Einstein tensor, $G_{\mu\nu}$,
has components $G_{00}=3H^2, ~G_{ii}=-a^2(2\dot{H}+3H^2)$.
Assuming that the stress-energy tensor $T_{\mu \nu}$ is described
by a perfect fluid of the form $T_{00}=\rho, T_{ii}=a^2p$, we find
\begin{equation}
\rho=\frac{3H^2}{\kappa^2}, \qquad
p=-\frac{2\dot{H}+3H^2}{\kappa^2}.
\end{equation}
We also immediately find from
equations~(\ref{gravi1})-(\ref{gravi2}) \bea \rho &=&
\frac{\gamma}{2} \dot{\sigma}^2 + V(\sigma)-24 H^3\dot{f},
\\
p &=&\frac{\gamma}{2} \dot{\sigma}^2-V(\sigma) +8 \frac{\dd}{\dd
t} (H^2\dot{f})+16 H^3 \dot{f}, \label{pressure} \eea where
$f=f(\sigma)$. Note that this differs from the usual expression,
due to the field-dependent coupling constant, $f(\sigma)$. To
guarantee that the energy density of the scalar field is always
positive, we require $\dot{f} H <0$.

We assume the scalar field obeys an equation of state, with an
equation of state parameter given by
\beq
w \equiv {p \over \rho}=-1-{2 \over 3}h ={2q-1 \over 3},
\eeq
where $q\equiv - a\ddot{a}/\dot{a}^2$ is the deceleration parameter. A
minimal criterion for getting an accelerating expansion is $\rho+3
p < 0$, which is clearly a violation of the strong energy
condition for some time-like vectors $\xi^\mu$; the latter states
that $T_{\mu\nu} \xi^\mu \xi^\nu \geq \frac{1}{2}
T_\lambda\,^\lambda \xi^\mu \xi^\nu$ or equivalently $\rho+3p\geq
0$ and $\rho+p\geq 0$. Note that
\begin{equation}
\rho+ 3p= 2\gamma\,\dot{\sigma}^2-2 V(\sigma) +24\frac{\dd}{\dd t}
(N H^2\dot{f}),
\end{equation}
where $N=\int H \,{\dd t}$, so that $\dd N/\dd t=H$. For $\dot{f}=
0$, the acceleration condition $\rho+3p <0$ holds when $V(\sigma)>
\gamma\,{\dot\sigma}^2$. In the case $\dot{f}\neq 0$, however,
whether the condition $V(\sigma)> \gamma \dot{\sigma}^2$ is
sufficient or not depends on the sign of the time-derivative of
the coupling, $f(\sigma)$. For a canonical scalar field (i.e.
$\gamma>0$), both $\dot{f}<0$ and $\dot{H}\leq 0$ hold, in
general. The acceleration condition $\rho+3 p<0$ is satisfied for
$\dot{f(\sigma)}<0$.
Similarly, we find,
\begin{equation}
\rho+ p= \gamma\,\dot{\sigma}^2+ 8 \frac{\dd}{\dd t}
\left(H^2\dot{f}\right) -8 H^3 \dot{f} >0,
\end{equation}
and not zero as it would be in a spacetime which is exactly de
Sitter: $\gamma=0$, $f(\sigma)=$ const. The null energy condition
$T_{\mu\nu}\chi^\mu\chi^\nu\geq 0$ (for some null vectors
$\chi^\mu$), or equivalently $p+\rho \geq 0$ may be violated by
allowing $\gamma<0$, or instead by taking
$\ln(H^2\dot{f})+\int\gamma\dot{\sigma}^2 {\dd t}<N $.

\section{General solutions}

To simplify the study of the model, we define the following
dimensionless variables:
\begin{equation}\label{def-variables}
x= \frac{\gamma\kappa^2}{2} \left(\frac{\dot\sigma}{H}\right)^2,
\quad
y= \frac{\kappa^2 V(\sigma)}{H^2},
\quad
u=8\kappa^2
f(\sigma) H^2,
\quad
h= \frac{\dot{H}}{H^2}.
\end{equation}
The equations of motion, (\ref{gravi1})-(\ref{scalar1}), then form
a set of second order differential equations (see the Appendix).
In the absence of the GB coupling, so $u=0$, we find a simple
relationship:
\begin{equation}
y=3+h, \qquad x=-h.
\end{equation}
A physically more intriguing case is $u\neq 0$, for which the
variables satisfy
\begin{eqnarray}
y&=&3+h+\frac{1}{2} \left[u^{\prime\prime}+ (5-h) u^\prime- 2
(h^2+h^\prime+5h) u\right],\label{sol-Y}
\\
x &=&-h-\frac{1}{2} \left[u^{\prime\prime}-(1+h) u^\prime- 2
(h^2+h^\prime-h) u\right],\label{sol-X}
\end{eqnarray}
where the prime denotes a derivative w.r.t logarithmic time or
number of e-folds $N\equiv \ln(a(t)/a_0)$. Thus there can exist a
large class of solutions with different $u(N)$.

The quantity $\Lambda(\sigma) \equiv V(\sigma)- f(\sigma) {\cal
G}$ may act as an effective potential for the field $\sigma$. It
is clear that $\Lambda(\sigma)$ is a second order differential
equation with non-constant coefficients: \bea \label{effCC}
\kappa^2\Lambda(\sigma) &=&H^2\left(y-3u(1+h)\right) \nn
\\
&=&\frac{H^2}{2} \left[ u^{\prime\prime} +\left(5-
h\right)u^\prime -2(8h+h^2+h^\prime+3)u\right] +H^2 (3+h)\,. \eea
When solving (\ref{effCC}) for $u$, one finds that the
\textit{homogeneous} part of the solution corresponds to solving
the \textit{complementary} differential equation, \beq
\label{homo-de}
u^{\prime\prime}+(5-h)u^\prime-2(8h+h^2+h^\prime+3)u=0. \eeq The
\textit{homogeneous} solution is the full solution when one makes
the ansatz \beq \kappa^2 \Lambda(\sigma) = H^2 \left( 3 + h
\right), \label{Vhomo} \eeq which removes all
\textit{nonhomogeneous} terms. Of course, $f(\sigma)=0$ (i.e.
$u=0$) is the trivial solution for (\ref{homo-de}), which
corresponds to the absence of Gauss-Bonnet coupling. Note that by
setting $\kappa^2 \Lambda = H^2 \left( 3 + h \right)$ one
effectively makes the ansatz, $V(\sigma)=f(\sigma)\,{\cal{G}} +
M_P^2 H^2(\sigma) \left( 3 + h \right)$, reducing the number of
arbitrary parameters of the model to one; fixing $h$ alone will
fix the function $u(\sigma(N))$, or vice versa. A salient feature
of our construction of cosmological solutions is that, while the
contributions coming from both the field potential $V(\sigma)$,
and the GB potential term, $V_{GB}(\sigma)$, may be large
separately, the effective potential, $\Lambda(\sigma)$, can be
exponentially close to zero at late times, as it dynamically
relaxes to a small value after a sufficiently large number of
e-folds of expansion.

A commonly discussed alternative is the following. One solves the
(modified) Einstein equations by making assumptions about the
(functional) form of the field potential $V(\sigma)$ as well as
the GB coupling $f(\sigma)$, which may be motivated by the leading
order terms obtained (see, for
example~\cite{Nojiri05b,Amendola:2005}), by a time-dependent
(cosmological) compactification of classical supergravities.
However, in this fashion, one may find a good approximation at
each energy scale, but the corresponding solution will have little
relevance when one attempts to study a wider range of scales,
including an inflationary era.

\subsection{Homogenous solution with constant $h$}

As a first reasonable approximation at low energy, we make the
ansatz $h\equiv \dot{H}/H^2=H^\prime/H \simeq$ const $<0$.
More specifically, when
$h=h_0$, the Hubble parameter is given by
\begin{equation}
H=\e^{\int h\,{\dd N}}=H_0\, \e^{h_0 N} \equiv \left( c_0-h_0 t\right)^{-1},
\end{equation}
where $H_0$ (or $c_0$) is an integration constant. Clearly, for
$t>0$, we require $h_0<0$ (and also $H_0>0$). The scale factor is
given by $a(t)= a_0 (c_0- h_0 t)^{-1/h_0}$, implying that the
universe accelerates when $-1<h_0<0$. For $t\simeq 0$ (and/or
$h_0\simeq 0$), the size of the universe in this regime grows
approximately as $\e^{H t}$, where $H\sim H_0\sim c_0^{-1}$.
Imposing (\ref{homo-de}), we find the
homogenous solution:
\begin{eqnarray}
u&=& u_1\,\e^{c_1 N} + u_2\,\e^{c_2 N} \label{uhomo}
\\
y&=& 3+h_0+3(1+h_0)\left(u_1\,\e^{c_1 N} + u_2\,\e^{c_2
N}\right)\,,
\\
x&=&-\frac{3}{2} \left(7+5h_0+\sqrt{9h_0^2+54h_0+49}\right)u_1
\,e^{c_1 N}  \nn
\\
&{}&-\,\frac{3}{2} \left(7+5h_0-\sqrt{9h_0^2+54h_0+49}\right)u_2\,
e^{c_2 N}-h_0 \,,
\end{eqnarray}
where \bea c_{1}=\frac{1}{2}\left(h_0-5+ \sqrt{9 h_0^2+54
h_0+49}\right)\,, \nn
\\
c_{2}=\frac{1}{2}\left(h_0-5- \sqrt{9 h_0^2+54 h_0+49}\right)\,,
\label{c1c2} \eea and $u_1,u_2$ are arbitrary at this stage.
Whether the coupling function $u$ is decreasing or increasing with
$N$ depends on the value of $h_0$. In particular, for $h_0\sim 0$,
we have $c_1\sim -6$ and $c_2\sim 1$, in which case the second
term on the rhs of (\ref{uhomo}) increases with $N$. To rescue it,
one might require to set $u_2\sim 0$. However, for $h_0\sim -1$
(and hence $w\sim -1/3$), since $c_1\to -2$ and $c_2\to -4$, only
the first term on rhs. of (\ref{uhomo}) is relevant for large N,
unless $u_2\gg u_1$. For purely exponential solutions we require
that $u_1$ and $u_2$ be real and $9 h_0^2+54 h_0+49\geq0$. This
means that either \beq h_0\leq-3-\frac{4\sqrt{2}}{3} \simeq -4.88
\quad \textrm{or} \quad h_0\geq-3+\frac{4\sqrt{2}}{3}\simeq -1.11,
\eeq implying that $w> 2.2$ or $w<-0.26$. However, only for
$h_0>-1$ (and hence $w<-1/3$), the universe is accelerating.

On the other hand, if $9 h_0^2+54 h_0+49<0$,
then the solution is written as
\begin{eqnarray}
u(N)&=& e^{(h_0-5)N/2}
\bigg[k_1\,\sin\left(N\sqrt{-(9 h_0^2+54 h_0+49)}\right)\nonumber \\
\qquad &{}& +\, k_2 \cos\left(N\sqrt{-(9 h_0^2+54
h_0+49)}\right)\bigg],
\end{eqnarray}
where $k_1, k_2$ are real. We see that in the latter case for
$h_0<5$ the GB contribution at late times (large $N$) becomes
vanishingly small. This is automatically satisfied as \beq -4.88 <
h_0 < -1.11, \eeq implying that $ -0.26<w<2.2$. Equations
(\ref{uhomo})-(\ref{c1c2}) provide, in some sense, a natural
solution for Einstein's equations, as the effective cosmological
constant, $\Lambda(\sigma)$, is dynamical and becomes arbitrarily
small with time, while simultaneously they provide the simplest
forms for the GB scalar coupling, $f(\sigma)$, the field
potential, $V(\sigma)$, and the kinetic energy, $K(\sigma)$.
However, this scenario cannot hope to effectively describe more
than one epoch in the history of the universe, as $h$ is not
dynamical.

\subsection{Homogenous solution with dynamical $h$}

Motivated by the form of the previous solution we now make no
assumptions about the form of the (slow-roll type) variable $h$,
but instead consider the ansatz that during a given epoch we may
make the approximation:
 \begin{equation}
 \label{sol-u-homo1}
u(\sigma)\equiv f(\sigma)\,H^2 =(\lambda-\xi(\sigma)) H^2 \approx
u_t\, \e^{\alpha_t N},
\end{equation} where the $time$, $t$, can
be {\it early} or {\it late}, where $|u_{\it early}|\gg |u_{\it
late}|$ and $|\alpha_{\it early}|> |\alpha_{\it late}|$. The
Gauss-Bonnet invariant, ${\cal G}=24 H^2 (H^2+\dot{H})=24
\left(\frac{\dot{a}}{a}\right)^2 \frac{\ddot{a}}{a}$, which is
positive for the accelerating solutions (i.e. $\ddot{a}>0$),
decays faster than the coupling $\xi(\sigma)$ blows up, but it may
well be that $H^2$ decays slower than $f(\sigma)$ (or
$\xi(\sigma)$). In the case $f(\sigma)\sim 1/H^2$, $\alpha\simeq
0$.

\begin{figure}[ht]
\begin{center}
\epsfig{figure=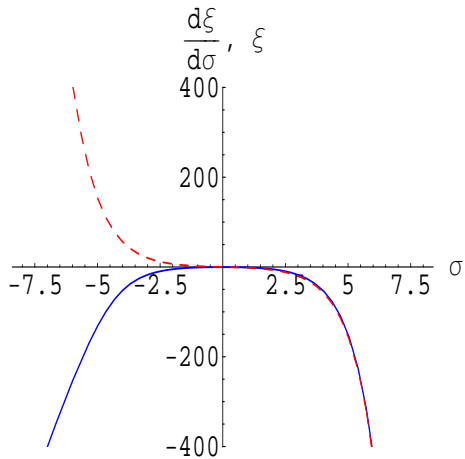,height=2.4in,width=2.8in}  \caption{
The function $\xi(\sigma)$ (solid line) is symmetric about
$\sigma\to -\sigma$, while its first derivative,
$\dd{\xi}/d\sigma$ (dotted line) is antisymmetric. Note that in
these plots we have set $\delta=1$.} \label{coupling-fig}
\end{center}
\end{figure}

One may motivate the above ansatz for the coupling $u(\sigma)$ in
two different ways. Firstly, the function $f(\sigma)$ as a
solution to a second order differential equation in $u(N)$ can
have two different branches, which may dominate at different time
scales, as can be seen above in the case $h\equiv
\dot{H}/H^2\simeq $ const. Secondly, in a known form for the
coupling $f(\sigma)$ ($\equiv \lambda-\xi(\sigma)$), derived from
heterotic string theory, the function $\xi(\sigma)$ may be given
by~\cite{Antoniadis93a,Easther:1996a}
\begin{equation}\label{xi-Easther}
\xi(\sigma)=\delta \ln[2\e^\sigma \eta^4(i\e^\sigma)],
\end{equation}
where $\delta$ is a constant and $\eta(b)$, with $b\equiv
i\e^{\sigma}$ is the Dedekind $\eta$-function which is defined by
$\eta(b)=\e^{i\pi b/12}\prod_{n=1}^{\infty}\left(1-\e^{2i\pi
{n}{b}}\right)$. As can be seen from the plots in
figure~\ref{coupling-fig}, $\xi(\sigma)/\delta$ can be well
approximated by $\ln {2}-\frac{2\pi}{3}\cosh(\sigma)$, while
$(\dd\xi/\dd\sigma)/\delta$ by $-\frac{2\pi}{3}\sinh |\sigma|$.

In the expansion of the function $\xi(\sigma)$ (and hence
$f(\sigma)$), the term proportional to $\e^{-\sigma}$ can dominate
at early times, $\sigma\ll 0$, while the term proportional to
$\e^{\sigma}$ can dominate at late times, $\sigma\gg 0$. The fact
that such an expansion is typical of string effective actions
represents, in our opinion, an interesting aspect of such models
and implies the existence of two periods of accelerating expansion
of the universe. Though the function $\xi(\sigma)$ is symmetric
about $\sigma\to -\sigma$, the coupling $u(\sigma)$ is not, since
the Hubble parameter $H$ is a monotonically decreasing function of
$N$~($\equiv\ln(a)$), or the field $\sigma$, and $H_{early} \sim
{10}^{23} {\rm eV} \gg H_{late} \sim {10}^{-33} {\rm eV}$.

By solving equation~(\ref{Vhomo}), or equivalently
equation~(\ref{homo-de}), we find
\begin{equation}
h(N)=-\hat{\beta}+\beta \tanh{\beta \Delta N}, \label{h-soln-main}
\end{equation}
where $\Delta N\equiv N-N_t$, with $N_t$ being an integration
constant, analogous to $\ln(a_0)$ in (\ref{defN}), and
\begin{equation}
\hat{\beta} =\frac{16+\alpha}{4},\quad
\beta=\frac{\sqrt{9\alpha^2+72\alpha+208}}{4}.
\end{equation}
Here we have suppressed the subscripts referring to the time on
$\alpha$. Because of our ansatz for $u(\sigma(N))$ of the
form~(\ref{sol-u-homo1}) the solution~(\ref{h-soln-main}) is a
good approximation at {\it early} or {\it late} epochs only. Of
course, one could make a more complicated ansatz for the function
$u(\sigma)$, motivated by some specific particle physics models,
and find the corresponding solution for $h(N)$. We believe that
the solution we have found above is sufficiently simple to explain
the evolution of the universe both at early time $N\gtrsim
N_{early}$ and at late times $N\gtrsim N_{late}$.

\begin{figure}[ht]
\begin{center}
\epsfig{figure=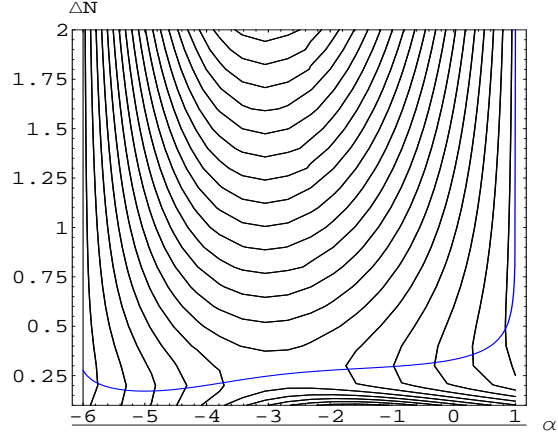,height=2.4in,width=2.9in}
\caption{ The contour plots of the effective potential
$\Lambda(\sigma)/H_0^2$ with the height in log scale. The single
solid (blue) line denotes where $\dd {\Lambda}/{\dd N}=0$, giving
the local maximum of the potential w.r.t. $\Delta N$. For larger
$\Delta N$, the potential generically decreases exponentially to
zero. } \label{con-V-fig}
\end{center}
\end{figure}

From~(\ref{h-soln-main}), we easily find that
\begin{equation}\label{Hubble-rate-gen}
H(N) = \e^{\int h(N) \dd N }= H_0\, \e^{-\,\hat{\beta} N}
\cosh{\beta \Delta N}, \end{equation} where, again, the time $t$
can be {\it early} or {\it late}. Writing the expressions for
$y(N)$ and $x(N)$ is straightforward, using
eqs.~(\ref{sol-Y})-(\ref{sol-X}). Analogous to an inflationary
type solution induced by a
conformal-anomaly~\cite{Starobinsky:1980te}, the solutions given
above are singularity-free. The effective potential therefore
takes the form
\begin{equation}
\Lambda(\sigma(N)) = \frac{H_0^2}{\kappa^2}
\left[3-\hat{\beta}+\beta \tanh{\beta \Delta N}\right]
\left(\cosh{\beta \Delta N}\right)^2\,\e^{-2\hat{\beta} N}.
\end{equation}
This possesses a local maximum around $0.15 \lesssim \Delta N
\lesssim 1$, depending upon the value of the slope parameter,
$\alpha$ (see figure~\ref{con-V-fig}) and generically decreases to
zero with increasing $N$. In fact, $\Lambda(\sigma)$ is almost
flat for $\Delta N>1$. As we are assuming that $N_{early}\ll
N_{late}$, the local maxima and `global minima' of the effective
potential can be at different energy scales for the early-time and
late-time universes. The kinetic energy, $K(\sigma)$, decreases
with $\Delta N$ for $\-6<\alpha<0$, and increases for
$0<\alpha<1$. The dimensionless variable, $x=K/H^2$, is
approximately constant with $\Delta N \gtrsim 10$ for a wide range
of $\alpha$.

\begin{figure}[ht]
\begin{center}
\epsfig{figure=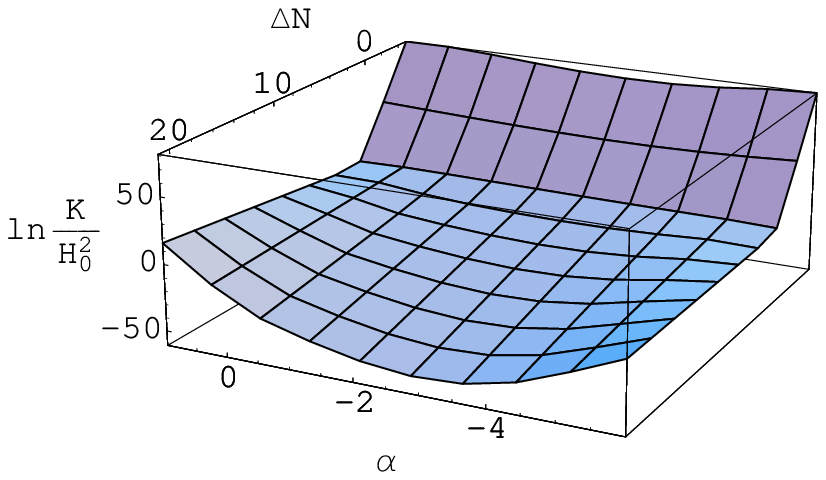,height=2.4in,width=2.9in}
\hskip0.1in
\epsfig{figure=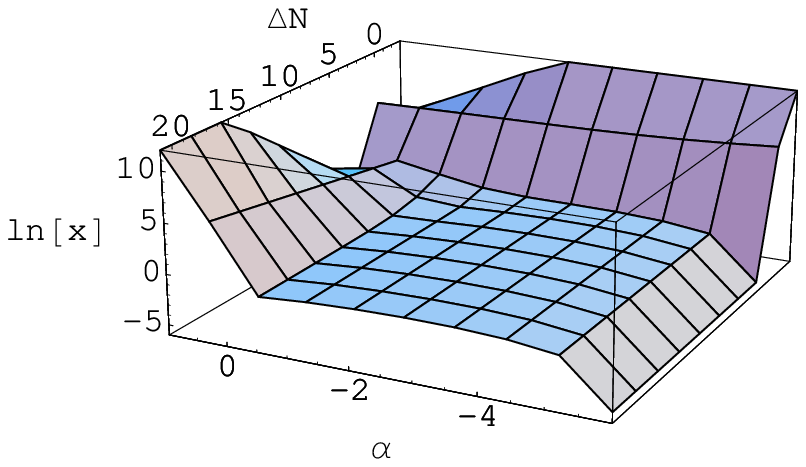,height=2.4in,width=2.9in}
\caption{ The kinetic term $K/H_0^2$ (left plot) and the
dimensionless variable $x$ (right plot), in logarithmic scales, as
functions of $\alpha$ and $\Delta N$, where $x \equiv (\gamma/2)
(\dot{\sigma}/H)^2 = K/H^2=(\gamma/2){\sigma^\prime}^2$. For
$\alpha> 1$ (or $\alpha < -6$), $K<0$ and hence $x<0$, leading to
a phantom type cosmology. For $-6< \alpha <0.45$, $K/H_0^2$
rapidly approaches zero. In all plots, where applicable, we have
set $u_t=10$. } \label{K-and-x-fig}
\end{center}
\end{figure}

One may express the field potential $V(\sigma)$ and the coupling
constant $f(\sigma)$ as functions of the field $\sigma$ itself. As
the second plot in figure~\ref{K-and-x-fig} shows, for
$\alpha\lesssim 0.45$, the parameter $x(N)$ is almost independent
of $N$, implying that $N= \lambda_0 \sigma $+ const; we will
choose this last constant to be $N_t$. For $\Delta N\gg 0$,
$\lambda_0~ (\equiv
\frac{1}{M_P}\sqrt{\frac{\gamma}{2x_0(\alpha)}})$ is a function of
the slope parameter $\alpha$ only. This leads to the following
expressions for the potentials:
\begin{eqnarray}
V(\sigma)&=& V_0\,\e^{-2\hat{\beta} \lambda_0 \sigma}
\left(\cosh(\beta
\lambda_0 \sigma)\right)^2\nonumber \\
&{}& \times
 \left[3-\hat{\beta}+ \beta \tanh (\beta\lambda_0\sigma) +3\hat{u_t}\left(1-\hat{\beta}+
 \beta \tanh (\beta \lambda_0\sigma)\right)\,\e^{\alpha(\lambda_0\sigma)}\right],\\
f(\sigma) &=& \frac{V_0}{H_0^4}
\,\hat{u_t}\,\e^{(2\hat{\beta}+\alpha) \lambda_0 \sigma}\,
\left({\rm sech}(\beta\lambda_0\sigma)\right)^2
  \e^{4\hat{\beta} N_t},
\end{eqnarray}
where $V_0\equiv M_P^2 H_0^2 \,\e^{-2\hat{\beta} N_t}$ and
$\hat{u_t} \equiv u_t\, \e^{\alpha N_t}$.

\begin{figure}[ht]
\begin{center}
\epsfig{figure=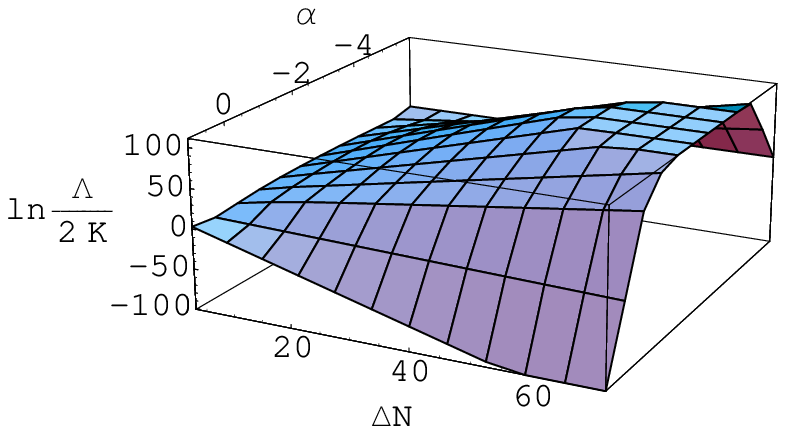,height=2.4in,width=2.9in}
\hskip0.1in
\epsfig{figure=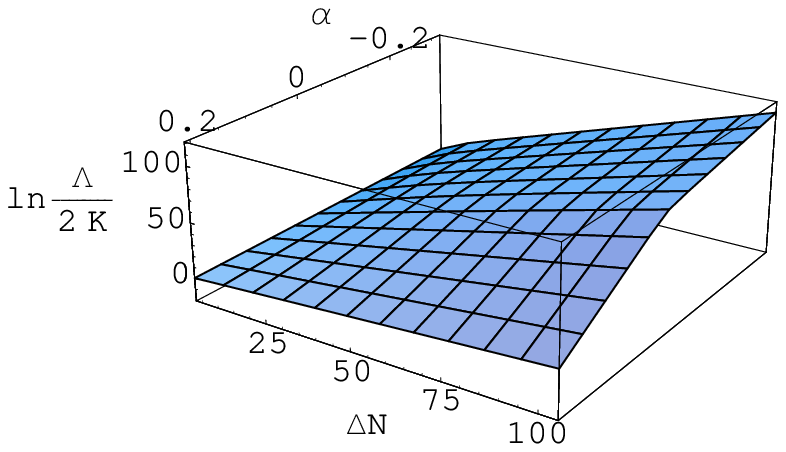,height=2.4in,width=2.9in}
\caption{ The ratio $\Lambda(\sigma)/2K(\sigma)$ in logarithmic
scale, in different ranges for $\alpha$. } \label{log-ratio-fig}
\end{center}
\end{figure}

For $-6< \alpha < 1$, so $\hat{\beta}> \beta$, the Hubble
parameter is a smoothly decreasing function of $N$ in this regime.
Typically, for $\alpha\simeq 1$ (or $\alpha \simeq -6$), since
$\hat{\beta}\simeq \beta$, we have $H\simeq H_0$, $h\simeq 0$ and
hence $w\simeq -1$: a stage where $\Lambda(\sigma)$ acts as a
cosmological constant. While, for $1>\alpha>-2$ (or $-6<\alpha
<-4$), we have $0> h > -1$ and hence $-1< w <-1/3$. When the value
of $\alpha$ is decreased from unity, the kinetic term lowers
towards zero at large $N$, while the parameter $x(N)$ is almost
constant (cf figure~\ref{K-and-x-fig}). For $\alpha\lesssim 0.28$,
the condition $\Lambda(\sigma)> 2K(\sigma)$ holds in general. But
in our model this does not necessarily mean the existence of an
accelerating phase; as the figure~\ref{log-ratio-fig} shows,
$\Lambda(\sigma)>2K(\sigma)$ holds even if $-2> \alpha >-4$; in
this last case we get $h< -1$ and hence $w>-1/3$, implying a
non-accelerating universe.

We will consider two epochs, an initial inflationary epoch where
$N$ is assumed to grow from an initial value $N_i \lesssim N_{\it
early}$, and a late-time deceleration/acceleration phase where $N$
becomes comparable to $N_{\it late}$, and $N \gtrsim N_{\it late}$
eventually holds. The universe starts to inflate when $N\gtrsim
N_{early}+0.5$. Consequently, at $N\gtrsim N_{early}$, we have a
stage of inflation. For $\beta(N-N_{early})\gtrsim 2.5$, the
scalar field begins to freeze in, such that $w< -1/3$; the actual
value of $w$ depends on the value of slope parameter $\alpha$,
see~\cite{IshBen05d}. After a certain number of e-folds, $\Delta
N= N_{f}-N_i$, our approximation, (\ref{sol-u-homo1}), with the
$time$ being $early$ breaks down. For $N \lesssim N_{late}$, the
universe is in a deceleration phase which implies that inflation
must have stopped during the intermediate epoch. Sometime later,
when $N> N_{late}$, subsequent evolution will be controlled by
(\ref{sol-u-homo1}) with the $time$ being $late$.


For $\beta \Delta N \gtrsim 2.5$, the variation in $h(N)$ is
small, so we can approximate it by
\begin{equation}\label{value-of-h}
h \simeq \left\{\begin{array}{l}
-\hat{\beta} +\beta  \quad (\Delta N\gg 0)\\
-\hat{\beta}- \beta \quad (\Delta N\ll 0).
\end{array} \right.
\end{equation}
Rearranging this approximation for $h$, we find
$\alpha=\{c_1,c_2\}$, as defined in (\ref{c1c2}). This tells us
that the earlier approximation, $h\approx h_0$, is a good `early
time' or `late time' approximation \textit{during} a given epoch.
Also note that for $h\leq 0$ (and hence $w \geq -1$) we require
$-6\leq \alpha \leq 1$. Picking $\alpha$ outside this range leads
to `big rip' type cosmologies for which $h>0$ (and hence $w<-1$)
in some regions of field space.

\subsection{Relaxation of dark energy}

There might be a large shift in the Hubble expansion rate during
inflation, viz $H_{before} \sim {10}^{23}\,eV \gg H_{after} \sim
{10}^{-33}\,eV$. One could therefore ask whether the solutions we
presented above explain a dynamical relaxation of vacuum energy
(or scalar potential) to the present value of dark energy, $\sim
{10}^{-120}\,M_P^4$, after a significant period of inflation. This
is quite plausible in our model. To quantify this, one considers
the ratio of the Hubble parameters before and after $N$ e-folds of
inflation, which is given by \beq \varepsilon=
\frac{\cosh\beta(\Delta N+N)\, \e^{-\hat{\beta}
N}}{\cosh\beta(\Delta N)}. \eeq As an illustration consider that
$\alpha\simeq 0.0143\simeq 1/70$, so that $q_{ini}\simeq 0^-$ at
$\Delta N \equiv N_{ini}-N_0 \simeq 0.33$. Assuming $70$ e-folds
of inflation, i.e. $N=70$, we find \bea \varepsilon \simeq
1.36\times {10}^{-12}. \eea This value represents a shift in
Hubble expansion rate, only during an accelerating epoch for which
the initial value of the deceleration parameter is zero. Indeed,
something like $70$ e-folds of expansion, as usually considered,
is the minimum for inflation, based on the assumption of a
constant Hubble rate. Practically, one requires much more
expansion than $e^{70}$ between the Planck time and the present.
As emphasized in~\cite{Liddle1994a}, one might need
$\tilde{N}\equiv \ln \frac{(a H)_f}{(aH)_i}\geq 70$, rather than
$N\equiv \ln ({a_f}{a_i})\geq 70$, to solve the various
cosmological conundrums, including the flatness and horizons
problems. The difference $N-\tilde{N}$, which is non-zero and
positive, as long as $-6<\alpha <1$, would characterize the extra
amount of expansion (e-folds) required by the decrease of $H$
during inflation.

In our model, the total number of e-folds $N$ required to get a
small ratio, like $\epsilon \sim {10}^{-56}$, depends on the value
of the expansion parameter $\alpha$, which may be related to the
slope of the potential coming purely from the Gauss-Bonnet
coupling, $V_{GB}(\sigma) \equiv f(\sigma)\,{\cal G} = 3 u(\sigma)
H^2(1+h)$. This value would be minimum, $N \sim 125$, for $\alpha
\gtrsim -2$ (or $\alpha \lesssim -4$), while it would be large,
$\Delta N\gtrsim {\cal O}(300)$, for $\alpha\lesssim 1$ (or
$\alpha\gtrsim -6$). In the former case, inflation could occur
slowly since $\frac{\dd}{\dd t} (\frac{1}{a H})\lesssim 0$, while,
in the later case, it might occur more rapidly since
$\frac{\dd}{\dd t} (\frac{1}{a H})\ll 0$. For example, for
$\alpha= 0$, the effective potential $\Lambda(\sigma) \sim
{10}^{-8} M_P^4$ decreases to the present value of dark energy,
namely, $\Lambda_0\sim {10}^{-120}\,M_P^4$, when $N_{total} \sim
326$.

Assuming the universe has undergone a sufficient number of e-folds
of expansion, like $N_{total} \gtrsim 125$, we find an effective
scalar potential that can dynamically relax its value to the
observed value of the cosmological constant, such that the field
$\sigma$ evolves towards its minimum and the equation of state
parameter falls in the range $-1 <w < -1/3$.

\subsection{Late-time acceleration}

The deceleration parameter $q$, which may be parameterized as a
function of red-shift factor $z$, is given by
$$ q(z)= -\frac{1}{H^2}\frac{\ddot{a}}{a}= -(1+h). $$
The value of the function, $h$, for the current epoch must be determined by
observation. For example, if the observed value of the deceleration parameter
$q(z)$ corresponds to $\simeq -0.6$, then \bea q_{{obs}} &=&
-1-h_{{obs}} \nn
\\
& \simeq & -1 + {\hat{\beta}}_{late}-\beta_{late} \simeq  -0.6\,,
\eea where we have made the late time approximation for
$h_{{obs}}$. Solving $h_{obs}=-0.4$ we find
\begin{equation}
\alpha_{late}\simeq -0.0148 \quad \mbox{or} \quad
\alpha_{late}\simeq -5.3851.
\end{equation}
Out of the two possible values of the slope parameter $\alpha$,
giving rise to the same value of deceleration parameter $q(z)$, we
should prefer to take the smaller value of $\alpha$, in terms of
its absolute magnitude, so that the coupling $u(\sigma(N))\sim
f(\sigma) H^2$ varies only slowly with the expansion of the
universe. In fact, the function $u(\sigma)$ is a measure of the
strength of scalar coupling with the curvature terms: $u(\sigma)$
is likely to depend upon the gauge coupling strength. And, in the
presence of matter sources, like standard model particles, the
running of the gauge coupling could be small if $|\alpha|\lesssim
{\cal O}(\frac{1}{N})$.

\begin{figure}[ht]
\begin{center}
\epsfig{figure=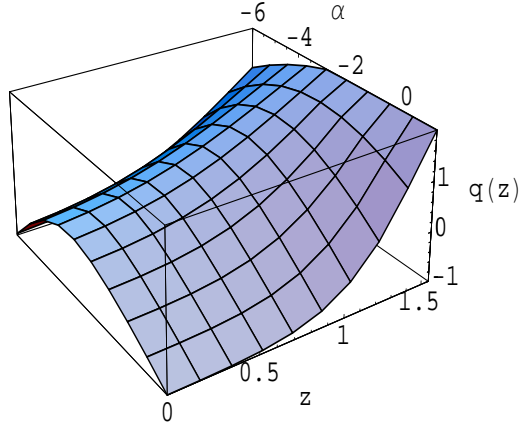,height=2.4in,width=2.8in}
\caption{The solution is modelled such that $\ln(1+z) \equiv
\Delta N=\ln\frac{a_f}{a_i}$. One can see a change from
deceleration to acceleration as the red-shift factor $z$ decreases
in the range $0.5\lesssim z \lesssim 1.5$.}
\label{red-shift-h-fig}
\end{center}
\end{figure}

Given the definition of redshift \beq 1+z=a_{now}/a_z \eeq we can
rewrite it as $1+z=e^{|\Delta N|}$, allowing us to plot recent
acceleration versus $z$. At late times, $N_0 = N_{late}$, we may
observe a significant variation in $h$ over the range of red-shift
$z<1.5$ (cf figure~\ref{red-shift-h-fig}), in excellent agreement
with observation \cite{Bennett03a}.

\subsection{Scalar potential: leading order behaviour}

It might well be that, asymptotically (i.e. at late times), both
the functions $h(N)$ and $u(\sigma(N))$ take (nearly) constant
values.  In this rather special limit
\begin{equation}
h(N) \rightarrow h_0<0, \quad u(\sigma(N)) \rightarrow u_0>
\frac{1}{h_0-1},
\end{equation}
we find the following solution
\begin{eqnarray}
H &=& \frac{1}{c_0-h_0 t}, \quad N = \int H\,{\dd t}, \\
{\sigma \kappa}&=& \pm \sqrt{\frac{2}{\gamma}} \sqrt{-h_0
(1+u_0-u_0 h_0)}\,N + const, \\
V(\sigma) &=& \frac{H_0^2}{\kappa^2}\, \e^{2 h_0 N}
\left(3+h_0-u_0 h_0 (5+h_0)\right) \equiv V_0
\,\e^{-\sigma/\sigma_0},
\end{eqnarray}
where $\sigma_0 \kappa =
\sqrt{\left(u_0-(1+u_0)/h_0\right)/2\gamma}$. It is not difficult
to see that, to leading order in $\sigma$, $f(\sigma)\propto
\e^{\sigma/\sigma_0}$. That is, the {\it ans\"atze} like
$V(\sigma)=V_0\,\e^{-\sigma/\sigma_0}$ and $f(\sigma)=f_0 \,
\e^{\sigma/\sigma_0}$ characterize only the late time attractors,
for which both $u(N)$ and $h(N)$ behave merely as constants.

Consider a more general case for which $h(N)\equiv
\dot{H}/H^2=h_0<0$, but $u(\sigma(N))=u_0\,\e^{\alpha N}$,
$\alpha\neq 0$. In this case, $\sigma$ is related to $N$ via
\begin{equation}
\sqrt{\frac{\gamma}{2}}\, {\sigma \kappa} =
\frac{\sqrt{c_1\,\e^{\alpha N}-4 h_0}}{\alpha}
-\frac{2\sqrt{-h_0}}{\alpha} \tanh^{-1}
\left(\frac{\sqrt{c_1\,\e^{\alpha N}-4 h_0}}{2\sqrt{-h_0}}\right),
\end{equation}
where
\begin{equation}
c_1\equiv 2 u_0 (\alpha-2h_0)(1-h_0-\alpha).
\end{equation}
Both signs of $\alpha$ may be allowed as long as $c_1> 4
h_0\,\e^{-\alpha N}$ holds. Indeed, $h(N)={H^\prime}/H \simeq 0 $
is special case, for which
\begin{equation}
{\sigma \kappa} = \sqrt{\frac{2}{\gamma}} \,\sqrt{\frac{2u_0
(1-\alpha)}{\alpha}}\,\e^{\alpha N/2}
\end{equation}
and hence $\Lambda(\sigma)= 3 M_P^2 H_{0}^2\equiv \Lambda_0$. Note
that, since the equation of state parameter $w=-1$,
$\Lambda(\sigma)$ acts as a cosmological constant term in the
regime $h\sim 0$. In this case $\sigma$ can be real even if
$\gamma<0$, provided that both $\alpha<0$ and $u_0>0$ hold
simultaneously. Nevertheless, one should be more interested in a
canonical scalar ($\gamma>0$), in which case one now requires
$0<\alpha<1$ and $u_0>0$. As we will see shortly, the slope
parameter $\alpha$ in this range possesses an interesting feature
that the spectrum of scalar (density) perturbation during
inflation is almost flat, giving a scale invariant power spectrum,
$n_s\simeq 1$.

\section{Inflation and cosmological perturbations}

It is generally believed that during inflation the inflaton and
graviton field undergo quantum-mechanical fluctuations, leading to
scalar (density) and tensor (gravity waves) fluctuations, which in
turn would give rise to significant effects on the large-scale
structure of the universe at the present epoch. The spectra of
perturbations may provide a potentially powerful test of the
inflationary hypothesis. One could ask whether the model presented
here will add to the search for a conceivable physical basis for
inflation. We believe this is possible.

\subsection{Slow roll variables}

In a standard scenario, one defines the slow roll variables, in
terms of the field potential $V(\sigma)$:
\begin{equation} \epsilon_v \equiv \frac{1}{2\kappa^2 {\sigma^{\prime}}^2}
\left(\frac{V^\prime}{V}\right)^2, \quad \eta_v \equiv
\frac{1}{\kappa^2 {\sigma^\prime}^2}\left(
\frac{V^{\prime\prime}}{V}-\frac{V^\prime}{V}
\frac{\sigma^{\prime\prime}}{\sigma^\prime}\right).
\end{equation}
As above, the prime denotes derivative with respect to $N$, not
the field $\sigma$. These definitions for slow roll parameters may
be justified since during the inflationary epoch $V(\sigma)\gg
V_{GB}(\sigma)$ holds. Note that $V_{GB}(\sigma)= f(\sigma) {\cal
G}= 3 u H^2 (1+h) \to 0$ as $h\to -1$. Because of the flatness of
$V(\sigma)$, $\sigma$ grows very slowly and essentially all the
inflation occurs when $V(\sigma) \gg V_{GB}(\sigma)$. Slow-roll
requires that $\vert \epsilon_v \,\vert \ll 1$, $\vert
\eta_v\,\vert \ll 1$.

To impose the conditions on slow roll variables in a physically
motivated and model-independent way, following
ref.~\cite{Liddle1994a}, we may define them in terms of the Hubble
parameter $H(\sigma)$ and its derivatives:
\begin{eqnarray}
\epsilon_H  &\equiv&  \frac{2}{\kappa^2}
\left(\frac{H_\sigma}{H}\right)^2=\frac{2}{\kappa^2
{\sigma^\prime}^2}
\left(\frac{H^\prime}{H}\right)^2, \\
\eta_H &\equiv & \frac{2}{\kappa^2} \frac{H_{\sigma\sigma}}{H}=
\frac{2}{\kappa^2 {\sigma^\prime}^2}
\left(\frac{H^{\prime\prime}}{H}-\frac{H^\prime}{H}
\frac{\sigma^{\prime\prime}}{\sigma^\prime}\right),
\end{eqnarray}
where, as before, primes denote derivatives w.r.t. $N$. One also
defines the following parameter, which is second order in
slow-roll expansion:
\begin{equation}
\xi_H \equiv \frac{1}{2\kappa^2} \left(\frac{H_\sigma
H_{\sigma\sigma\sigma}} {H^2}\right)^{1/2}=\left(\epsilon_H \eta_H
-\sqrt{2\epsilon_H}\,
\frac{\eta_H^\prime}{\sigma^\prime}\right)^{1/2}.
\end{equation}
(The $\xi_H$ defined above is not to be confused with the coupling
function $\xi(\sigma)$ we defined before; here we are adopting the
notations which are standard in literature and
we will not refer to $\xi(\sigma)$ in this section). In fact, these
definitions of slow roll variables may be of wider applicability
than those defined in terms of $V(\sigma)$, as they are based on
the fact that inflation occurs as long as $\frac{\dd}{\dd t}
(\frac{1}{a H})< 0$ holds. That is, during inflation, the comoving
Hubble radius, $1/(aH)$, must necessarily decrease, so that
physical scales can grow more rapidly than the Hubble radius.
Inflation ends at $\frac{\dd}{\dd t} \left(\frac{1}{k}\right)= 0$,
where $k\equiv a H \equiv a_{e}\,H(\sigma)\,\e^{-\Delta N}$ is, by
definition, a scale matching condition, where $a_e$ is the value
of the scale factor at the end of inflation, at which a coming
mode crossed outside the (cosmological) horizon.

Let us analyze the last case considered above in some detail.
First, note that, defining $\sigma \to \sigma/\sqrt{\gamma}$, we
can always absorb the coupling constant $\gamma$ into the
slow-roll parameters. Without loss of generality, henceforth we
define $\epsilon\equiv \epsilon_H/\gamma$, $\eta\equiv
\eta_H/\gamma$, $\xi\equiv \xi_H/\gamma$; similar arguments would
apply to other quantities, like, scalar and tensor indices,
defined below.

At the level of approximation we are considering in this section,
namely, $V(\sigma)\gg V_{GB}(\sigma)$, the tilt parameter, up to
first order terms ($\epsilon$ and $\eta$), may be given by
\begin{equation}
n_s \simeq 1-4 \epsilon + 2 \eta.
\end{equation}
(We will discuss below about the validity of such a relation in
the presence of the Gauss-Bonnet term.) As is generally the case,
a value $n_s<1$ is easier to produce than $n_s>1$ for the model we
are considering here; $n_s\sim 1$ is the value that makes the
(scalar) perturbation small in all scales, see, e.g.
\cite{Liddle:1993fq}, for a review. In our model, $n_s>1$ is
possible only if one allows $\alpha>1$ (or $<-6$). This last
demand is however not a physically motivated one since the kinetic
energy of the field $\sigma$ is negative in at least some regions
of field space.
\begin{figure}[ht]
\begin{center}
\epsfig{figure=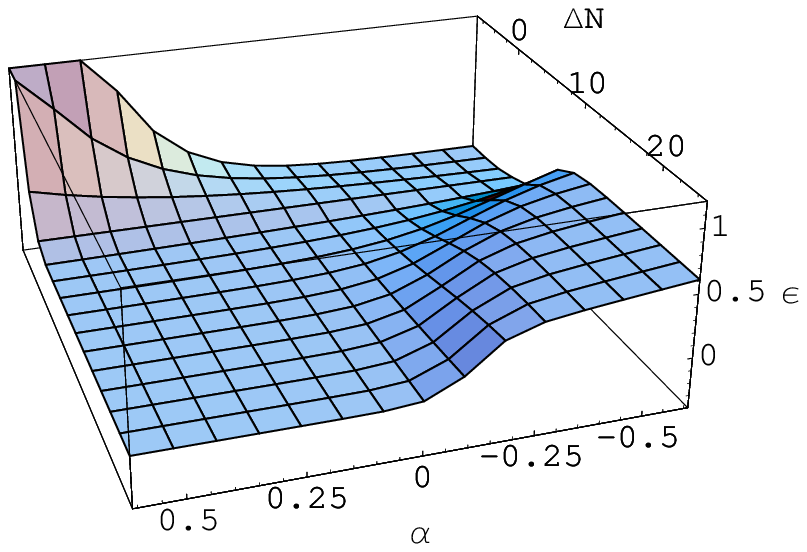,height=2.4in,width=2.8in}
\hskip0.1in
\epsfig{figure=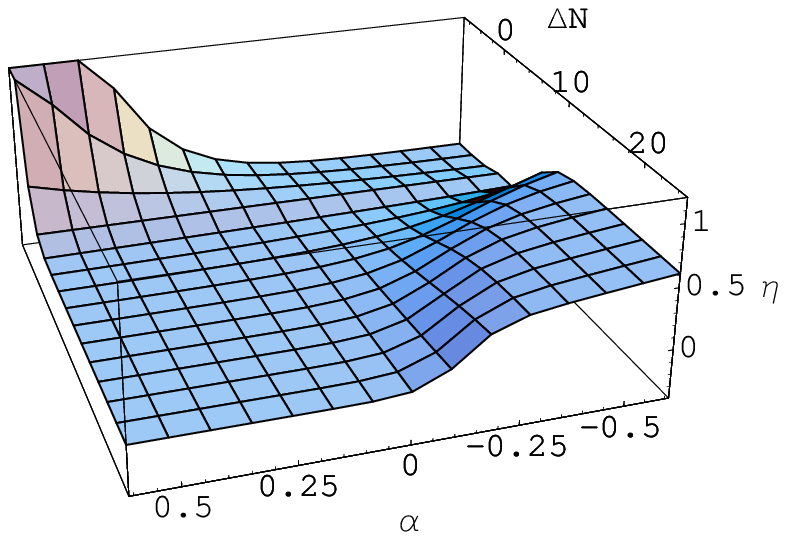,height=2.4in,width=2.8in}
\caption{ The slow roll variables $\epsilon$ and $\eta$ as
functions of $\alpha$ and $\Delta N$. For a small and positive
$\alpha$, the universe is decelerating before inflation, $\Delta
N<0$, but $\epsilon$ (and also $\eta$) quickly takes a small value
($\epsilon, \eta \ll 1$) for $\Delta N\gtrsim 10$ and
$\alpha>-0.2$, leading to inflation. In fact, $\epsilon\sim 0$
also for $\alpha\simeq -6 $. Outside the range $-6\leq \alpha \leq
1$, we get $\epsilon<0$, leading to a phantom type cosmology. The
condition for acceleration to occur is $\epsilon< 1$, so the
universe is not accelerating for $-4 < \alpha <-2$.}
\label{epsilon-eta-fig}
\end{center}
\end{figure}
\begin{figure}[ht]
\begin{center}
\epsfig{figure=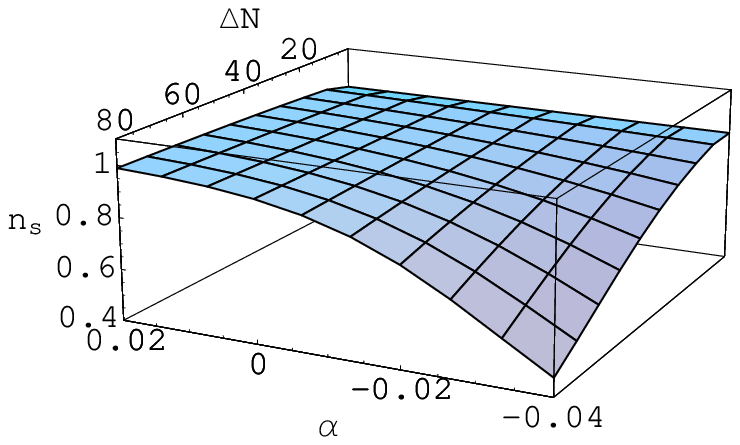,height=2.2in,width=2.6in}
\epsfig{figure=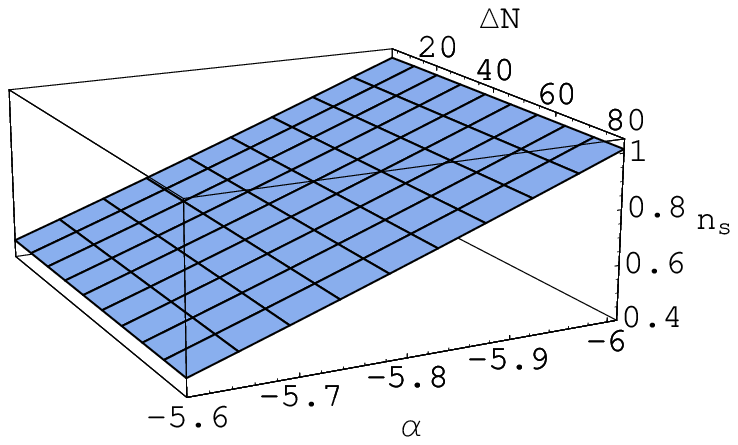,height=2.2in,width=2.6in}
\caption{ The spectral index $n_s$ as a function of slope
parameter $\alpha$ and $\Delta N$. Both for $\alpha\gtrsim 0$ and
$\alpha\ll 0$, the spectral index is independent of $\Delta N$.}
\label{s-index-Hubble-fig1}
\end{center}
\end{figure}
\begin{figure}[ht]
\begin{center}
\epsfig{figure=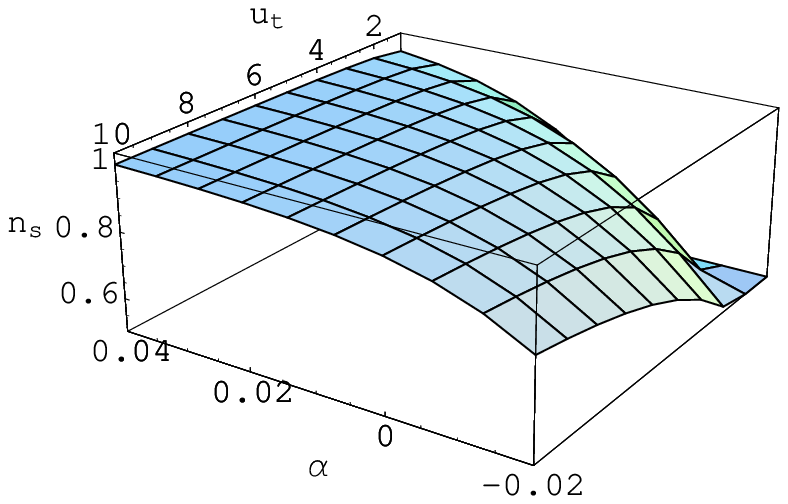,height=2.4in,width=2.8in}
\hskip0.1in
\epsfig{figure=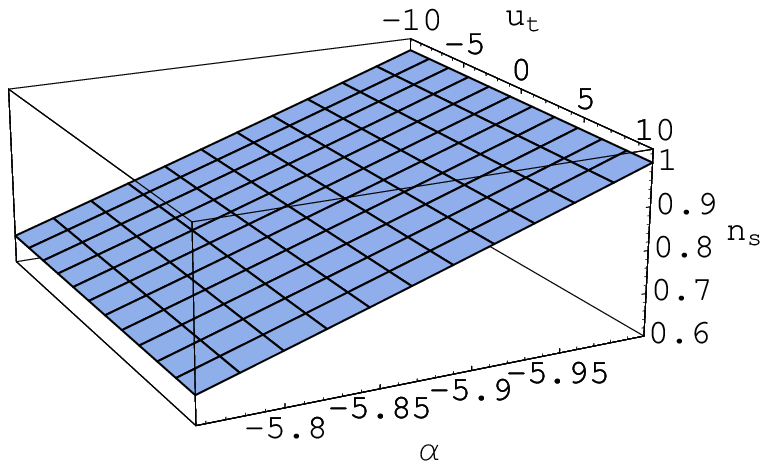,height=2.4in,width=2.8in}
\caption{ The spectral index $n_s$ as a function of $\alpha$ and
$u_t$, with a fixed value of $\Delta N=70$. The plots do not
change much while including terms second order in slow-roll, viz,
$n_s \simeq 1-4\epsilon+2\eta -2(1+c)\epsilon^2
-\frac{1}{2}(3-5c)\epsilon \eta+\frac{1}{2}(3-c)\xi^2$, where
$c\simeq 5.08$, as defined in~\cite{Liddle1994a}. A small
difference is that now a smaller value of $u_t$ is required so as
to get the same value of $n_s$.} \label{s-index-Hubble-fig2}
\end{center}
\end{figure}
\begin{figure}[ht]
\begin{center}
\epsfig{figure=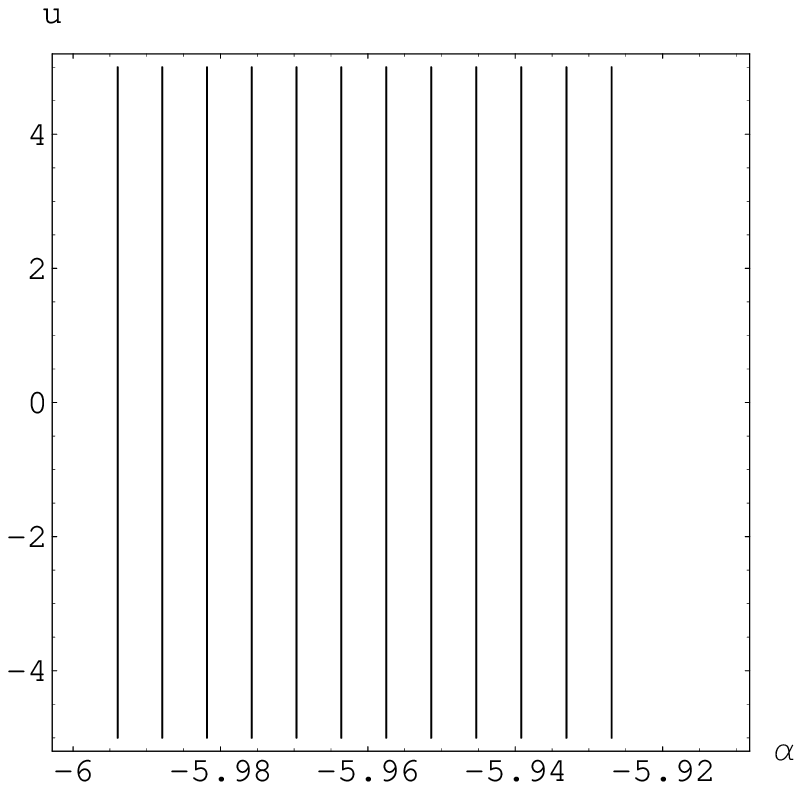,height=2.5in,width=2.8in}
\hskip0.3in
\epsfig{figure=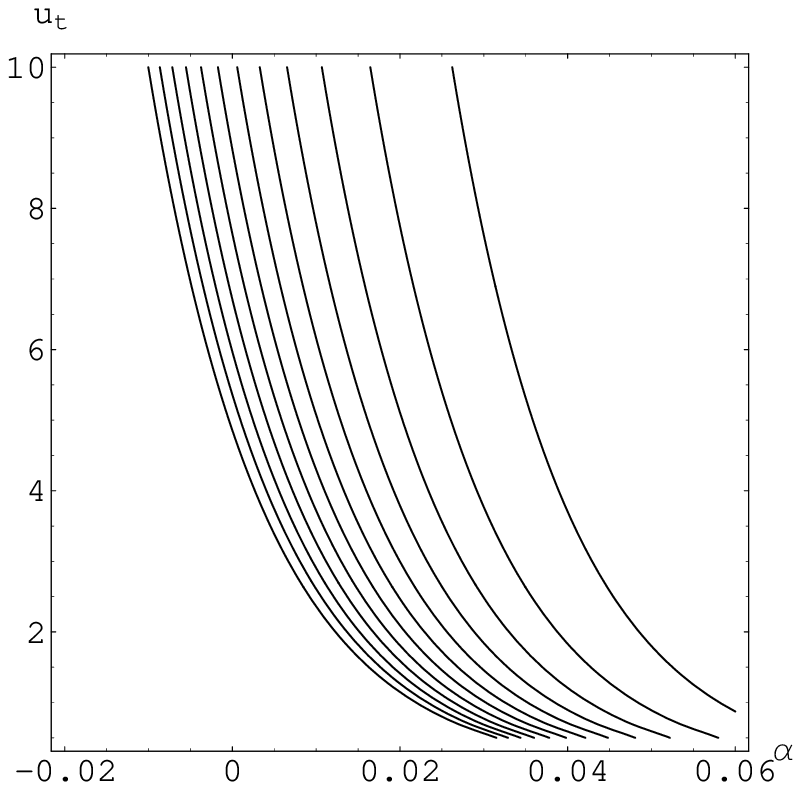,height=2.5in,width=2.8in}
\caption{ The contour plots of $u_t$ vs $\alpha$ in the range
$n_s=[0.89, 1]$, with a fixed value of $\Delta N=70$. The spectral
index, $n_s$, decreases (increases) from left to right in the left
(right) plot; $n_s>1$ can be obtained only for $\alpha<-6$ (or
$\alpha>1$). For $\alpha>0$, more positive is the value of
$\alpha$, smaller will be the coefficient $u_t$ giving rise to the
same value of $n_s$. For $\alpha<< 0$, however, $n_s$ is
insensitive to the value of $u_t$. For $\alpha>0$, the signs of
running of spectral index can be different between the $u_t<0$ and
$u_t>0$ cases. } \label{cont-s-index-fig}
\end{center}
\end{figure}

In our model, the slow roll variables $\epsilon$ and $\eta$ (and
also $\xi$) can vary with both the exponent and coefficient of the
coupling constant $u(\sigma(N))$, i.e. $\alpha$ and $u_t$, in
addition to the number of e-folds, $N_f-N_i=\Delta N$ (cf
figure~\ref{epsilon-eta-fig}). In turn, as the figures
\ref{s-index-Hubble-fig1} and \ref{s-index-Hubble-fig2} show, it
is possible to get a value of $n_s$ close to unity in a wide range
of $u_t$ (or $\Delta N$) by suitably choosing $\alpha$.
Interestingly, as shown in figure~\ref{cont-s-index-fig}, a value
of $n_s$ in the range $[0.89,1]$ may easily be obtained by taking
different combinations of $\alpha$ and $u_t$. One obtains the
value $n_s\sim 0.95$, in excellent agreement with recent
observational data from WMAP, by taking
\begin{equation}
\Delta N \sim 70, \quad \alpha_{early} \sim - 0.01, \quad
u_{early} \sim 22.
\end{equation}
For $\alpha<0$, with a smaller value of $\Delta N$, one also
requires a smaller value of $u_{t}$; when $\Delta N= 50$ and
$\alpha= -0.01$, we find $n_s\simeq 0.97$ for $u_{early}\simeq
31$, while $n_s\simeq 0.95$ for $u_{early}\simeq 18$. Of course, a
positive value for the slope parameter, $\alpha$, is also allowed,
as long as $\alpha<1$ holds. Note, for $\alpha>0$ and $N>0$ the
function $u(\sigma(N))\sim \e^{\alpha N}$ grows with proper time,
$t$, or logarithmic time $N$. But, in our construction, any
contribution like this, coming from Gauss-Bonnet term is exactly
cancelled with the \textit{homogeneous} part (i.e. terms
multiplied by $u(\sigma)$ and its derivatives) of the field
potential $V(\sigma)$. Also, both the Hubble parameter $H$ and
$u(\sigma) H^2$ can be slowly decreasing functions of the number
of e-folds, $N$, or the field $\sigma$.

Note that if $V(\sigma)=0$, then Einstein gravity may not be an
effective theory at low energy, since the term $u(\sigma) H^2$ can
easily dominate the Einstein-Hilbert term, $R/\kappa^2$, which is
proportional to $3 H^2$. Even though the time scale for such
effect to occur can be extremely large, given that $\alpha$ is
very close to zero (i.e. $f(\sigma)$ scales nearly as
$1/H(\sigma)^2$) and the coefficient $u_t$ can be extremely small,
this provides an additional justification for considering an
effective action with a non-trivial field potential, $V(\sigma)$.

\subsection{Generation of perturbations}

Note that for the solution (\ref{Hubble-rate-gen}) the scale
factor of the universe evolves as
\begin{equation}\label{scale-fac-gen}
a(t) \sim (t+t_0)^{1/m}\sim |\tau|^{1/(m-1)},
\end{equation}
where $m \equiv \hat{\beta}-\beta<1$ and $\tau$ is the conformal
time. In the case of power-law inflation, such as this, the
amplitudes for scalar and tensor fluctuations may be given
by~\cite{Stewart:1993a,Lidsey:1995a}
\begin{eqnarray}\label{sca-ten-ampl}
A_s &= & \frac{4 A_0^{(s)}}{5 M_P^2} \left(1+ 0.46 \epsilon-0.73
\eta\right)
\frac{H \sigma^\prime}{|\,h |}, \\
A_T &= & \frac{2 A_0^{(T)} H}{5\sqrt{\pi} M_P}
\left(1-0.27\epsilon\right).
\end{eqnarray}
Two remarks are in order. First, for each of these perturbations
the value of $H$ (and hence $h$) and
$\sigma^\prime~(=\frac{\dd\sigma}{dN})$ must be evaluated when the
wavelength of the perturbation becomes of the order $H^{-1}$.
Practically, it is more convenient to specify them as functions of
e-folds $N$ before the end of inflation. Second, the above
expressions, with $A_0^{(s)}=A_0^{(T)}= 1$, best approximate the
results in a model with a self interaction potential $V(\sigma)$
alone~\cite{Stewart:1993a}, i.e. without the GB coupling. In fact,
in the $f(\sigma)\neq 0$ case, $A_0$'s are generally functions of
$f(\sigma)$ and $H$, not unity. As long as the GB term is only
sub-leading to the scalar potential $V(\sigma)$, the above results
would be available to leading order.

Following~\cite{Hwang97,Tsujikawa:2002qc}, one may calculate the
coefficients related to scalar and tensor type perturbations
\begin{equation}
A_0^{(s)}= \frac{{s_{(s)}}^{-\nu/2}}{\sqrt{Q_{(s)}}}, \quad
A_0^{(T)}= \frac{{s_{(T)}}^{-\nu/2}}{\sqrt{Q_{(T)}}},
\end{equation}
where $\nu \equiv \frac{3}{2}+ \frac{m}{1-m}$ and
\begin{eqnarray}
s_{(s)} &\equiv & 1+
\frac{4F^2}{1+2F}\left(2h+\frac{F-\ddot{f}}{1+2F}\right) \left(2
x+
\frac{6 F^2}{1+2 F}\right)^{-1},\\
\sqrt{Q_{(s)}} &\equiv & \frac{1}{\kappa}
\bigg|\frac{H}{\dot{\sigma}}\bigg| \left(2 x+ \frac{6 F^2}{1+2
F}\right)^{1/2}
\left(\frac{1+2 F}{1+3 F}\right),\\
Q_{(T)} &=& 1+2 F,\\
s_{(T)} Q_{(T)} &=& 1+2\ddot{f},
\end{eqnarray}
where $F \equiv \kappa^2 \dot{f}H$ and $x\equiv \frac{\gamma}{2}
\kappa^2 \left(\frac{\dot{\sigma}}{H}\right)^2$. In the limit $F
\to 0$ (or $\dot{f} \to 0$), one recovers the standard result for
which $s_{(s)}= s_{(T)}=1$, $\sqrt{Q_{(s)}}= \sqrt{\gamma}$ and
$\sqrt{Q_{(T)}}=1$. As can be seen at the level of the field
equations, equations~(\ref{main1})-(\ref{main3}), the condition
$0< |F|\ll 1$, or alternatively
\begin{equation}
F  \equiv \kappa^2 \dot{f} H = \frac{H^2}{8}
\left(\frac{u}{H^2}\right)^\prime = \frac{1}{8} (u^\prime-2h u)\ll
1,
\end{equation}
is equivalent to $V(\sigma)\gg V_{GB}(\sigma)$. This may be
satisfied, for example, by choosing $u_t=10$, $\Delta N=70$ and
$\alpha< 0$. $|F| < 1$ implies that $\dot{\sigma} H$ decays faster
than the function $\frac{\dd\xi}{\dd{\sigma}}$ grows. Since
$\ddot{f}=-\ddot{\xi}
=-{\dot\sigma}^2\frac{\dd^2\xi}{\dd\sigma^2}-{\ddot\sigma}
\frac{\dd\xi}{\dd\sigma}$, calculating the above quantities
requires knowledge about how the scalar field $\sigma$ varies with
time, which is model dependent.

\begin{figure}[ht]
\begin{center}
\epsfig{figure=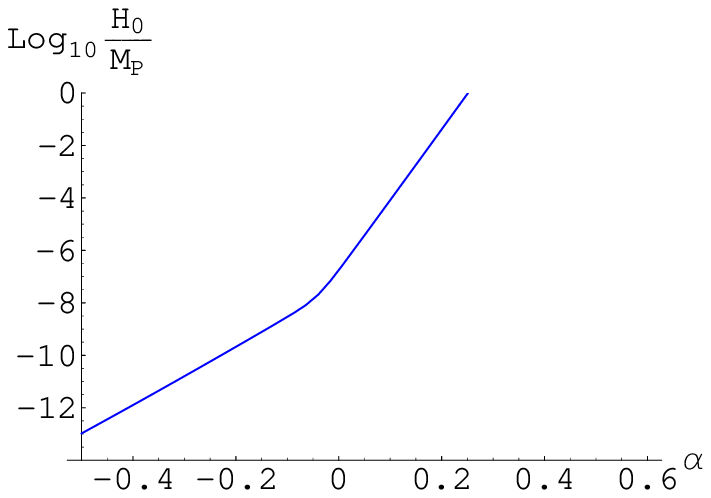,height=2.4in,width=2.8in} \hskip0.1in
\epsfig{figure=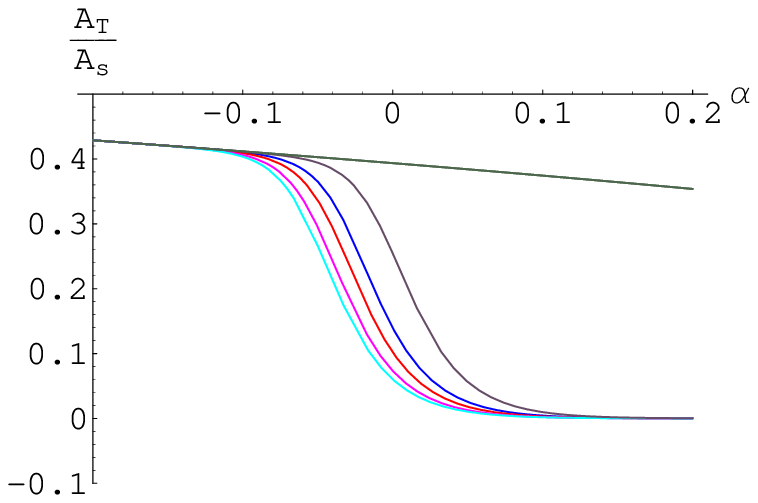,height=2.4in,width=2.8in}
\caption{(a) (left plot) The logarithm of the scale $H_0/M_P$ vs
$\alpha$, giving rise to the value $A_S= 2\times {10}^{-5}$, which
may be matched to the density contrast, $\delta_H$, at
Hubble-radius crossing. A particular value of $\alpha$ fixes the
energy scale $H_0$ in terms of Planck mass, e.g., for
$\alpha\simeq 0.0284$, $H_0\simeq 10^{-6}\,M_P$. (b) (right plot)
The tensor-to-scalar ratio, $A_T/A_S$, as a function of $\alpha$:
from left to right $u_t=30,20,10,5,1$. This ratio is independent
of the number of e-folds. The single (horizontal) line corresponds
to $u_t=0$, in which case $A_T/A_S$ (and hence $n_T$) can be large
($r\lesssim 0.43$). } \label{scalar-tensor-fig2}
\end{center}
\end{figure}

\begin{figure}[ht]
\begin{center}
\epsfig{figure=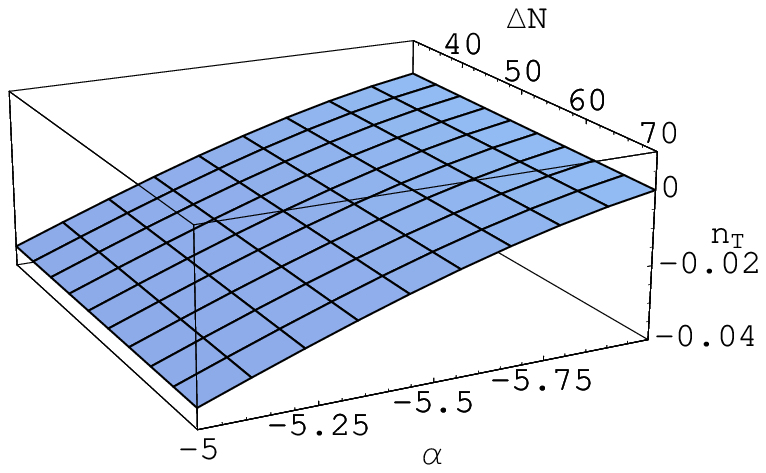,height=2.4in,width=2.8in}
\hskip0.1in
\epsfig{figure=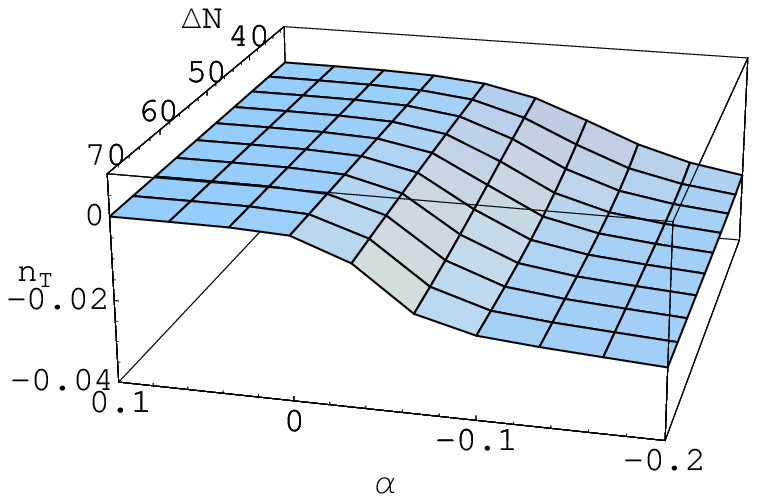,height=2.4in,width=2.8in}
\caption{ The tensor index $A_T$ as a function of $\alpha$ and
$u_t$, with fixed $\Delta N=70$. The tensor modes are almost
negligible, $n_T\simeq 0$, for $\alpha\gtrsim 0$ (or $\alpha\simeq
-6$), or equivalently, for $\epsilon\ll 1$.}
\label{At-As-ratio.fig}
\end{center}
\end{figure}

Since $h\leq 0$, a small negative value of $F$ would normally
decrease the value of $s_{(s)}$ but would increase the value of
$\sqrt{Q_{(s)}}$. This is opposite for the tensor modes: $s_{(T)}$
will increase but $\sqrt{Q_{(T)}}$ will decrease. Although $A_s$
and $A_T$ both vary with $\Delta N$, or the choice of scale
$k\equiv aH$ at which a comoving mode crossed outside the horizon,
the ratio $A_T/A_S$ is essentially independent of such a scale, or
the number of e-folds, $\Delta N=N_f-N_i$. The tensor spectral
index may therefore be given
by~\cite{Copeland:1993a,Lidsey:1995a},
\begin{equation}
n_T \simeq -2 \frac{A_T^2}{A_S^2} \left[2-n_s-
\frac{A_T^2}{A_S^2}\right].
\end{equation}
In the plots of figures \ref{scalar-tensor-fig2} and
\ref{At-As-ratio.fig}, one may find a significant deviation for $F
\sim -0.5$. Interestingly, for the solutions that we found above,
this last last condition is extremely difficult to satisfy since
it requires $u^\prime-2h u \simeq -4$, but $u^\prime-2h u \simeq
0_-$ at the end of inflation for $\alpha<0$, and $u^\prime-2h u >
0$ for $\alpha>0$.

Let us discuss about the validity of the relation, $n_s-1\simeq
-4\epsilon+2\eta$, in the presence of a non-trivial GB coupling.
As shown in~\cite{Hwang97gf}, in general terms, the spectral
indices of the scalar and tensor-type perturbations may be given
by
\begin{equation}
n_s-1 = 3- 2\nu_s, \quad n_T=3-2 \nu_t.
\end{equation}
In our case, since $\nu_s=\frac{3}{2}+\frac{m}{1-m}$, it is
sufficiently clear that $n_s-1= -\frac{2m}{1-m}\le 0$. For the
tree-level solutions found by Gasperini and
Veneziano~\cite{Gasperini:1994}, the authors of~\cite{Hwang97}
found the unpleasing result $n_s\simeq 4$ and $n_T\simeq 3$. The
solutions given in~\cite{Gasperini:1994} are characterized by the
additional presence of a dilaton background $\phi(t)$, with
$V(\phi)=0$. This suggests, in our case, that the $V(\sigma)=0$
case is not illuminating at least in view of inflationary
paradigm. Indeed, the solutions found in~\cite{Gasperini:1994},
namely,
\begin{equation}
a(\tau)\propto |\tau|^{(1-\sqrt{3})/2}, \quad \phi(\tau)=-\sqrt{3}
\ln|\tau|
\end{equation}
are non-accelerating, so it is inconsistent in using
non-inflationary solutions to estimate the inflationary
parameters, like $n_s$.


In our model, it is quite possible that during an inflationary
epoch the field $\sigma$ changes only slowly, so the quantity
$x(N)\sim (\dot{\sigma}/H)^2=(\sigma^\prime)^2$ remains almost
constant over exponentially large range of wavelengths, leading to
an almost flat spectrum of perturbations of metric. Fluctuation of
the field $\sigma$ leads to the effect that the duration of
inflation is increased due to a local delay of time near the exit
from inflation, which may be given by
\begin{equation}
\delta t = \frac{\delta\sigma}{\dot{\sigma}}\sim \frac{1}{2\pi
\sigma^\prime}.
\end{equation}
This may lead to a local density increase such
that\cite{Mukhanov:1990me}
\begin{equation}
\delta_H \sim \frac{\delta\rho}{\rho} \sim
\frac{H}{2\pi\,\sigma^\prime}.
\end{equation}
Inflation also leads to creation of perturbations of the field
$\sigma$ with wavelength greater than $H^{-1}$. The average
amplitude of scalar perturbations during a typical time interval
$H^{-1}$ is given by~\cite{Linde:1982}
\begin{equation} |\delta \sigma
(x) \,|=\frac{H}{2\pi}.\end{equation} Indeed, this result, in its
more rigorous form $\frac{\dd}{\dd t}\langle \sigma^2\rangle
=\frac{H^3}{4\pi^2}$, was independently obtained
in~\cite{Starobinsky82a}. A useful observational constraint is the
following. If $\sigma$ changed very slowly during inflation, then
$H/\sigma^\prime$ remained almost constant over exponentially
large range of wavelengths. The cosmic microwave background
constraint is such that~\cite{Bunn:1996a}
\begin{equation}
\delta_H \sim 1.92\times {10}^{-5}.
\end{equation}
If this quantity is to be matched (precisely) with the
scalar-index $A_S$, as the results given in~\cite{Liddle:1993fq}
suggest, then, in our model, we can reproduce $\delta_H$ of this
magnitude by taking, for instance, $H_0\sim {10}^{-7} M_P$ and
$\alpha\sim -0.01$.

\section{Towards reheating in an inflationary universe}

It is essential to have a good reheating process after inflation
in a cosmological model. Recall that in our model the time
evolution of $\sigma$ is given by (cf equation~(\ref{scalar1}))
\begin{equation}\label{evo-equation}
\frac{\dd}{\dd
t}\left(\frac{\gamma}{2}\,\dot{\sigma}^2+\Lambda(\sigma) \right)
=- 6H \left(\frac{\gamma}{2}\dot{\sigma}^2\right)-\delta.
\end{equation}
where $\Lambda(\sigma)\equiv V(\sigma)-f(\sigma){\cal G}$ and
$\delta\equiv f(\sigma) \frac{\dd {\cal G}}{\dd t}$. Note that the
effective potential for the dynamics of the scalar field is
$\Lambda(\sigma)$ rather than $V(\sigma)$. The first term on the
r.h.s. represents the energy loss caused by the expansion of the
universe and the $\delta$ term represents the energy density per
unit time which is drained from the field $\sigma$ though
time-variation of the Gauss-Bonnet term. Physically, such a term
is expected since the modulus field $\sigma$, which obtains a
non-zero mass due to its vacuum expectation value, $\langle \sigma
\rangle$, is coupled to the curvature tensor.

Suppose that, initially, $\dot{\sigma}\sim 0$ and ${\dd
\Lambda}/{\dd t} \sim 0$ ($\dd\Lambda/\dd\sigma$ is not
essentially zero there). The evolution
equation~(\ref{evo-equation}) then implies that $\delta\sim 0$ and
hence
\begin{equation}
H\sim H_0 \left(1+ \e^{4(N_0-N)}\right)^{1/4} \quad \Rightarrow
\quad h \sim -\frac{\e^{4(N_0-N)}}{1+ \e^{4(N_0-N)}},
\end{equation}
where $H_0$ and $N_0$ are some integration constants. Suppose we
start inflation at a point where $N< N_0$~\footnote{One can define
the scale factor as $a\equiv \e^{\omega(t)}$, so that $N\equiv
\ln(a(t))=\omega(t)$ and $N_0=\omega_0$. } and $h\equiv
{\dot{H}}/{H^2}=H^\prime/H \simeq - 1$, so that the field
equations satisfy
\begin{equation}
3\dot{\sigma} H  \sim -\frac{1}{4} \left(\frac{\dd f}{\dd
\sigma}\right)^{-1}, \quad \ddot{\sigma} \sim 12 H \dot{\sigma}^3
\frac{\dd^2 f}{\dd \sigma^2} \ll 3 H\dot{\sigma}.
\end{equation}
For $N > N_0$, the universe experiences a nearly constant Hubble
flow $H\sim H_0$, leading to an early inflation (exponential
expansion) of the universe, i.e., $a \sim \e^{H_0\, t}$. The
physical Hubble radius $H^{-1}$ increases with the number of
e-folds $N$ as the field $\sigma$ rolls down its potential and
gets out of its local extremum (a point of inflection or critical
point). Inflation would be eternal (i.e. without an exit) if
$\delta\simeq 0$ holds for all times. However, this is generally
not the case since the Hubble flow is damped by an adequate cosmic
friction term, and most of the evolution of the universe would be
described by the $\delta\neq 0$ or $\frac{\dd
f(\sigma)}{\dd\sigma}\neq 0$ solution. At this point, one also
notes that the term $\delta$ can change its sign between
accelerating ($\frac{\ddot{a}}{a}>0$) and decelerating
($\frac{\ddot{a}}{a}<0$) solutions, since ${\cal G}=24
\left(\frac{\dot{a}}{a}\right)^2 \frac{\ddot{a}}{a}$. That is,
given that $f(\sigma)$ does not change its sign at the transition
point, $\ddot{a}=0$, the term $\delta$ acts as a (positive)
friction term during an accelerating phase
($\frac{\ddot{a}}{a}>0$), while it is opposite during a
decelerating phase.

So far we have neglected the couplings of the scalar field to
radiation (matter) fields. In the case the effective potential
$\Lambda(\sigma)$ possesses a local minimum, such couplings may
cause the rapid oscillatory phase to produce particles, leading to
{\it
reheating}~\cite{Albrecht82a,Kofman97a,Kofman94a}~\footnote{However,
in non-oscillatory models, for example a model of quintessence
characterized by a single exponential potential, the instant
preheating mechanism proposed by Felder et al~\cite{Felder98}
might be more efficient for particle production, which works even
for a potential without a minimum.}. To this end, we can introduce
some matter fields, which constitute ordinary matter and radiation
field released by the decay of the field $\sigma$. For a
homogeneous cosmological model with vanishing space curvature, we
find
\begin{eqnarray}
\left(\frac{\dot{a}}{a}\right)^2 &=& \frac{\kappa^2}{3} (\rho_\sigma+\rho_r),\\
\frac{\dd }{\dd t}\left(\frac{\gamma}{2}\dot{\sigma}^2
+\Lambda(\sigma)\right)&=&
-6\frac{\dot{a}}{a}\,\left(\frac{\gamma}{2}\dot{\sigma}^2\right)
-\delta,\label{drag-term}
\end{eqnarray}
where $\rho_\sigma= \frac{\gamma}{2}\dot{\sigma}^2 + V(\sigma) -24
\dot{\sigma} H^3 \frac{d f(\sigma)}{d\sigma}$. The $\delta$ term
in (\ref{drag-term}) resembles a drag term, which transfers energy
from the motion of $\sigma$ and dumps it in the form of a
radiation background. The equation for the evolution of the energy
density in radiation (particle) is given by
\begin{equation}
\dot{\rho_r}+4H \rho_r -\delta =0.
\end{equation}
Once reheating is completed the universe enters a standard
radiation dominated FRW phase:
\begin{equation}
p_0 =\frac{1}{3} \rho_0=\frac{1}{3} \rho_r, \quad \rho_0\sim
\frac{1}{a^4}, \quad H\sim \frac{1}{a^2} \sim \frac{1}{t}.
\end{equation}
The field $\sigma$ thereafter remains subdominant for most of the
time and only at late times, $N\gtrsim N_{late}$, when the
potential becomes sufficiently shallow, does one get acceleration,
$w< -1/3$. This would then make the model complete.

\begin{figure}[ht]
\begin{center}
\epsfig{figure=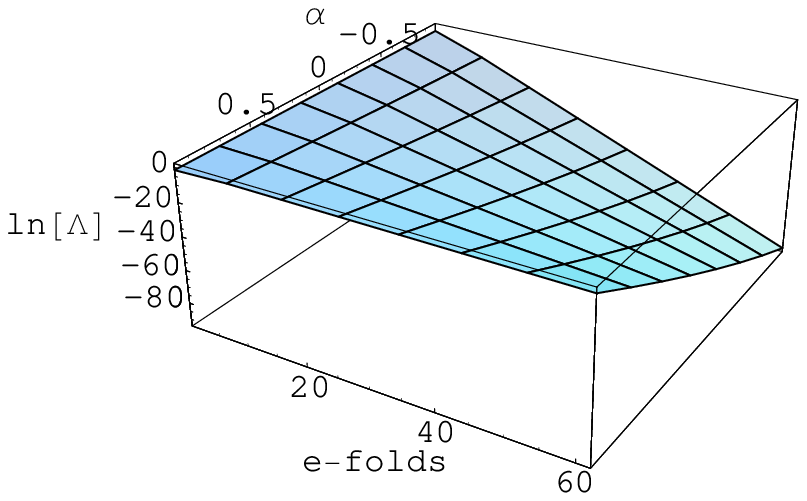,height=2.4in,width=2.9in}
\hskip0.05in
\epsfig{figure=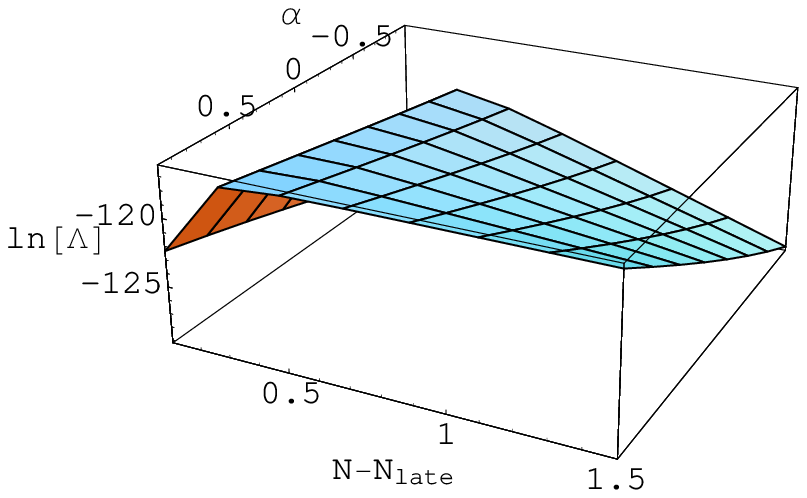,height=2.4in,width=2.9in}
\caption{Approximate behaviours of the effective potential
$\Lambda(\sigma)$ (not to scale) at early (left plot) and late
(right plot) times; we have chosen the free (integration)
parameters $H_{0, \rm early}$ and $H_{0, \rm late}$ such that the
plots mimic a realistic feature that $\Lambda(\sigma)_{\rm
early}\gg \Lambda(\sigma)_{\rm late}$. }
\label{scalar-poten-both.fig}
\end{center}
\end{figure}

A few remarks are in order. The first is related to a reheating
process after inflation, the details of which depend upon a choice
of the potential $V(\sigma)$ and the coupling $f(\sigma)$. More
generally, it is essential to know whether or not the effective
potential $\Lambda(\sigma)$ has a minimum. So far in this section
we have not specified the form of $\Lambda(\sigma)$; our
discussion of particle production above is rather qualitative.

Note that for the solution given in subsection (4.2),
$\Lambda(\sigma)$ does not have a local minimum, rather it has a
local maximum (cf figure~\ref{scalar-poten-both.fig}). In this
sense the conventional reheating process may not work and an
alternative reheating method need to be employed: an efficient
method of reheating is the {\it instant preheating} proposed by
Felder et al~\cite{Felder98}. In this case the field $\sigma$
decays when it rolls down the potential, thereby producing heavy
particles. However, details of the reheating process may be
somewhat different and complicated in our model for at least two
reasons. Firstly, as compared to a standard model, there is an
extra friction-like term, namely $\delta$, which is non-zero as
long as $\frac{\dd f(\sigma)}{\dd\sigma}\neq 0$. Secondly, in our
ansatz for $u(\sigma)$ ($\equiv 8\kappa^2 f(\sigma)H^2$), we
assumed that $u(\sigma)\propto \e^{\alpha_t N}$, where the time
$t$ can be {\it early} or {\it late}. In order to retrieve the
full potential, we may have to sew these two potentials in some
way, possibly creating a barrier (and hence a minimum) somewhere
between them. Work in this direction is currently underway.
Numerical studies show that for an ansatz of the type $u(\sigma) =
u_{\rm early}\,\e^{\alpha_t N}+ u_0(N)+ u_{\rm
late}\,\e^{\alpha_{late} N}$, where $u_0(N)\sim {\cal O}(1)$,
$\Lambda(\sigma)$ can have an effective local minimum, where
inflation could end. This is in accordance with our observation in
subsection (4.2), that inflation must have stopped during the
intermediate phase. In fact, some of our solutions presented in
the Appendix~\footnote{In the Appendix we do not make any specific
ansatz for $u(\sigma)$ but instead allow some of the variables to
take their limiting values, like $x\simeq $ const and $h\simeq $
const} possess an effective local minimum. In any case, the {\it
instant preheating} proposed in~\cite{Felder98} is perhaps the
most efficient method of particle production in our model, as it
works even for a potential without a minimum.

The second issue is related to the nucleosynthesis bound. As is
known, for inflation driven by a scalar field, there exists a
tight constraint on the allowed magnitude of $\Omega_\sigma \equiv
\frac{\rho_\sigma}{\rho_\sigma+\rho_m}< 0.1-0.2$ at the time of
nucleosyntheis; see, e.g., \cite{Ferreira97a}. This often places a
constraint on the model parameters, for instance on the slope of
the (effective) potential. One could ask whether or not the
nucleosynthesis bound will not be violated in our model. To
address this question properly, we would need to know the precise
form of the (effective) potential.

It is always possible to constrain some of the parameters in our
model by allowing $V(\sigma)$ and $f(\sigma)$ to take some
specific (functional) form. In subsection (4.2), we found a
solution by making a specific ansatz for $f(\sigma)$ (or
$f(\sigma) H^2$). For this solution, when the gravity is coupled
to matter and radiation, the necleosynthesis bound, in terms of
the allowed magnitude of $\Omega_\sigma$, constrains the values of
the parameters $\beta$ and $\hat{\beta}$, namely
$4|\hat{\beta}-\beta|^2 \lambda_0^2
> \frac{n}{\Omega_\sigma^{\rm max}}$, where $n=3$ ($4$) for matter
(radiation) and $\lambda_0\equiv \frac{1}{M_P}
\sqrt{\frac{\gamma}{2x_0(\alpha)}}=|\sigma^\prime|\Z{\beta\Delta
N>2}$. The nucleosysthesis bound may not be violated for
$(\hat{\beta}-\beta)\lambda_0 \gtrsim \sqrt{5}$, given that
$\Omega_\sigma^{\rm max}\lesssim 0.2$; in arriving at this result
we have assumed that the effective potential after a certain
number of e-folds ($\beta \Delta N > 2$) is approximated by
$\Lambda(\sigma) \propto
\e^{-2(\hat{\beta}-\beta)\lambda_0\sigma}$, since
$\tanh(\beta\Delta N)\to 1$. Also note that in this limit,
$V(\sigma)\propto \e^{-2(\hat{\beta}-\beta)\lambda_0\sigma}$, but
with a different proportionality constant.

An inflationary type potential that we arrived at from studying
symmetries of the field equations is approximated by
$V(\sigma)=V_0 \exp[-p(\sigma) \kappa\sigma]$, with $p$ being a
function of the field $\sigma$ or the number of e-folds $N$. Thus,
the reheating process discussed in \cite{Tashiro} may be useful in
our model. Finally, we would like to note that our model might
inherit some of the features of {\it quintessential inflation}
proposed by Peebles and Vilenkin~\cite{Peebles:1998qn} in which
the potential consists of two parts: $\lambda(\sigma^4+ M^4)$
($\sigma <0$) for inflation and $\lambda M^8/(\sigma^4+M^4)$
($\sigma \geq 0$) for quintessence; both models can lead to
tracker solutions at late times, though the forms of scalar
potentials are quite different.


\section{Conclusions}

In this paper we presented an analysis of
accelerating/inflationary cosmologies by introducing in the
effective action a field dependent Gauss-Bonnet coupling, other
than a standard field potential for the field $\sigma$. We find
that the dark energy hypothesis fits into a low energy
gravitational action where a scalar field is coupled to the
curvature squared terms in Gauss-Bonnet combination. It is
established that a GB scalar-coupling can play an important and
interesting role in explaining both the early and late-time
evolutions of the universe as well as providing a mechanism for
reheating.

That we are able to explain accelerating universes using exact
cosmological solutions in a modified Gauss-Bonnet theory, leading
to a small deviation from the $w=-1$ prediction of non-evolving
dark energy (or a cosmological constant) is likely to have a
serious impact in search of a viable dark energy model. Our work
also provides extension of quintessence (or time-varying
$\Lambda$) model in which part of the dark energy comes from a
field dependent Gauss-Bonnet interaction term.

One of the key results is this: in the absence of a Gauss-Bonnet
coupling, the tensor/scalar ratio is usually non-zero. However,
with a non-trivial scalar Gauss-Bonnet coupling, i.e.,
$f(\sigma)\neq $, or effectively, $u(\sigma) \equiv f(\sigma) H^2
\sim \e^{\alpha N} \neq 0$, such a ratio can be negligibly small
if the expansion parameter $\alpha$ takes a small positive value,
$\alpha \gtrsim 0.1$, and hence $n_s\lesssim 1$ and $n_T\simeq 0$,
leading to Harrison-Zel'dovich spectrum. Unlike a naive
expectation, the inclusion of a (scalar) field dependent
Gauss-Bonnet coupling $f(\sigma)$, in addition to a field
potential $V(\sigma)$, into the effective action, could make the
observability of tensor/scalar ratio and related inflationary
parameters more achievable.

We emphasize that, in contrast to previous analysis, our
calculations have all been implemented by the functional forms of
the scalar potential and GB scalar coupling, as suggested by the
symmetry of the field equations, rather than choosing particular
model dependant forms for them. We have given in the Appendix the
exact solutions for some special cases, about which a general
comparison can be made in terms of the homogeneous solutions we
presented in the bulk part of the paper.

Regardless of whether the model studied here appears natural or
otherwise, it should be observation that determines whether or not
it is correct. The current and future observations might make
stronger demand on theoretical precision of inflationary
parameters, including, the scalar and tensor spectral indices, and
are certain to constrain a number of parameters of our model
tightly, including the Gauss-Bonnet coupling constants.

\section*{Acknowledgements}

This work was supported in part by the Marsden fund of the Royal
Society of New Zealand. We wish to thank M Sami, Ewan Stewart and
David Wiltshire for discussions and helpful remarks.

\section{Appendix}
\renewcommand{\theequation}{A.\arabic{equation}}
\setcounter{equation}{0}

In terms of the dimensionless variables defined in
equation~(\ref{def-variables}), the equations of motion,
(\ref{gravi1})-(\ref{scalar1}), form a set of second order
differential equations:
\begin{eqnarray}
0&=&- {3}+x+y -3 (u^\prime-2 h u), \label{main1}
\\
0&=& u^{\prime\prime}- 2 h^\prime u -h u^\prime +2 u^\prime -4 h u
- 2 h^2 u + x-y+2h+3, \label{main2}
\\
0&=&x^\prime + 2(h+3) x + y^\prime +2 h y - 3 (h+1)(u^\prime-2 h
u). \label{main3}
\end{eqnarray}
where \beq X^\prime \equiv \frac{d X}{d N}= a \frac{d X}{d
a}=\frac{1}{H}\frac{dX}{dt}\,, \eeq so that \beq N=\int H\,
{dt}=\ln \left(\frac{a(t)}{a_0}\right) \label{defN} \eeq measures
the logarithmic expansion of the universe.

Note, the logarithmic time, or the number of e-folds, $N$, is a
monotonically increasing function of proper time $t$. However, its
sign depends on the assumption of what the scale $a_0$ represents.
If $a_0$ is the initial value of $a(t)$, such that $a(t)\geq a_0$,
then $N$ starts from zero and take a large positive value when
$a(t)\gg a_0$. However, if one wants $a_0$ to represent the
present value of the scalar factor, then $N$ usually starts from a
large negative number when $a(t)\ll a_0$ and becomes positive only
when $a(t)> a_0$.

Here we would like to present some exact solutions for some
special cases, about which a general expression can be obtained
for various parameters of the model, like, the field potential
$V(\sigma)$ and the coupling constant $f(\sigma)$, in terms of the
scalar field $\sigma$.

\subsection{$x(N)=x_0$ and $h(N)=h_0$}

As a reasonable approximation, at late times, consider that the
kinetic term $K(\sigma)\propto H^2(\sigma)$, so
$\dot{\sigma}/H\simeq$ const. Specifically, when $x \equiv
\kappa^2 \frac{K(\sigma)}{H^2(\sigma)}=x_0$, the field equations
reduce to
\begin{eqnarray}
&& 0=-3+x_0+y - 3(u^\prime-2 h u), \nonumber \\
&& 0=u^{\prime\prime}-(h+1) u^\prime +2h(1+u-hz)-2 u h^\prime+2
x_0.
\end{eqnarray}
$h\equiv \dot{H}/H^2$ is a kind of slow-roll variable, and thus
may be treated as a constant in at least some region of field
space. This is the case, for example, for power-law inflation,
namely, $a(t)\propto t^{1/m}$ where $m<1$. For $x= x_0$ and $h =
h_0$, we find the following interesting solution:
\begin{eqnarray}
u(N)&=& u_0+ u_1\, \e^{2h_0 N}+ u_2\, \e^{(1-h_0)N}, \\
y(N) &=& y_0+ y_1\, \e^{(1-h_0)N},
\end{eqnarray}
where $y_0=3+h_0(1-5 u_0 - h_0 u_0)$, $y_1 \equiv 3 u_2 (1-3
h_0)$, and $u_0, u_1, u_2$ are the integration constants. We also
find that
\begin{equation}\label{sol-sigma-xh}
\kappa\sigma = \pm N \sqrt{\frac{2 x_0}{\gamma}}+ const, \quad
x_0\equiv -h_0(1+u_0-h_0 u_0).
\end{equation}
This further implies that
\begin{eqnarray}
V(\sigma)& =& \frac{H_0^2}{\kappa^2} \left(y_0\,\e^{2h_0 N}
+ y_1\,\e^{(1+h_0)N}\right),\\
V_{GB}(\sigma) &=& \frac{3 H_0^2}{\kappa^2} (1+h_0)
\left(u_0\,\e^{2h_0 N}+ u_1\,\e^{4h_0 N} +
u_2\,\e^{(1+h_0)N}\right).
\end{eqnarray}
Using (\ref{sol-sigma-xh}), one may express these potentials as
functions of the field $\sigma$ itself, namely
\begin{eqnarray}
V(\sigma) & \sim & V_0\,\e^{-\sigma/\sigma_0}+ V_1\,
\e^{\frac{h_0+1}{2h_0} (\sigma/\sigma_0)},\\
f(\sigma) &\sim & f_0\,\e^{\sigma/\sigma_0}+ f_1 + f_2\,
\e^{\frac{3h_0-1}{2h_0} (\sigma/\sigma_0)},
\end{eqnarray}
where $\mp \kappa \sigma_0 h_0 \equiv \sqrt{\frac{x_0}{2\gamma}}
>0$. By combining the above expressions, we find
\begin{equation}
\Lambda(\sigma) = \frac{H_0^2}{\kappa^2} \left(A\,\e^{2h_0 N}+B\,
\e^{4 h_0 N}+C\, \e^{(1+h_0) N}\right),
\end{equation}
where $A \equiv 3(1-u_0)+h_0(1-8u_0-u_0 h_0)$, $B\equiv
-3(1+h_0)u_1$ and $C\equiv -12 h_0 u_2$. Equivalently,
\begin{equation}
\Lambda(\sigma) \equiv  \Lambda_0\,\e^{-\sigma/\sigma_0}
+\Lambda_1\,\e^{-2\sigma/\sigma_0}+ \Lambda_2
\,\e^{-\frac{1+h_0}{2h_0} (\sigma/\sigma_0)}.
\end{equation}
Thus, since $h_0 \leq 0$, it is the term proportional to $\e^{4
h_0N}$ or $\e^{(1+h_0)N}$ which is of greater interest at early
($N \lesssim 0$) or late ($N \gg 0$) times. In particular, when
$h_0\sim 0$, we find
\begin{equation}
V(\sigma)= 3 M_P^2 H_0^2 (1+u_2\,\e^N), \quad V_{GB}(\sigma) = 3
M_P^2 H_0^2 (u_0+u_1+u_2\,\e^N).
\end{equation}
In this case, since $\Lambda(\sigma)\sim 3 M_P^2 H_0^2
(1-u_0-u_1)$, $\Lambda(\sigma)$ acts as a cosmological constant
term, which is positive for $u_0+u_1<1$.

\subsection{$u(N)=u_0$ and $x(N)=x_0$}

Let us recall that in the original action
\begin{eqnarray}
 \frac{R}{2\kappa^2}&=& \frac{3H^2}{\kappa^2} \left(2+h\right),\\
f(\sigma) {\cal G} &=& 24 f(\sigma) H^4 \left(1+h\right) =
\frac{3H^2}{\kappa^2} \left(1+h\right) u(\sigma).
\end{eqnarray}
In a general situation, the coupling constant $f(\sigma)$
increases with logarithmic time, $N$, while the Hubble expansion
rate, $H(N)$, decreases with $N$. To this end, let us consider the
case where the coupling constant $f(\sigma)$ scales as $1/H^2$, so
that the function $u(\sigma)$ is constant, i.e. $u(\sigma) \simeq
$ const $\equiv u_0$. In general, we would require $u_0<
\frac{2+h}{1+h}$, because any contribution coming from the higher
order curvature corrections in low energy string effective actions
should be, at least in a low energy scale, an order of magnitude
smaller than the contribution from the Einstein-Hilbert term.
Also, the function $u(\sigma(N))$, which is likely to depend upon
the gauge coupling strength, has to be an extremely slow varying
function of proper time so that it assumes a nearly constant
value.

Let us further demand that the kinetic term $K(\sigma)$ scales as
$H^2$, so that $x(N)=\kappa^2 \frac{K}{H^2}\equiv x_0$. In this
case, the explicit solution is given by
\begin{equation}
y(N)= 3-x_0-6 u_0 h(N), \quad h(N)= h_1 +\beta \tanh\beta
{(N-N_0)},
\end{equation}
where
\begin{equation}
h_1\equiv \frac{u_0+1}{2 u_0}, \quad \beta\equiv
\frac{\sqrt{(u_0+1)^2+4 x_0 u_0}}{2u_0}.
\end{equation}
Further, a simple calculation shows that
\begin{eqnarray}
\sigma &= & \pm \frac{\sqrt{2}}{\kappa}\sqrt{\frac{x_0}{\gamma}} N
+
const,\label{sol-sigma-uh}\\
 H&=&H_0 \e^{h_1 N} \cosh \beta {(N-N_0)},
\end{eqnarray}
where 
$N_0$ is arbitrary. For a canonical scalar, $\gamma>0$, we have
$x_0 \ge 0$. A smaller value of $x_0$ makes the potential flatter
and hence increases the period of inflation.

Using (\ref{sol-sigma-xh}), or equivalently $N\equiv
\sqrt{\frac{\gamma}{2 x_0}}\, \sigma \kappa +$ const (we will
choose this last constant to be $N_0$), we find
\begin{eqnarray}
V(\sigma)&=& H_0^2 M_P^2 \,\e^{\frac{u_0+1}{u_0}
\sqrt{\frac{\gamma}{2 x_0}}\sigma\kappa } \cosh^2\beta
\left(\sqrt{\frac{\gamma}{2 x_0}}
\sigma\kappa\right)
\nonumber
\\
&{}& \quad \times \left[-x_0-3u_0-6 u_0\beta \tanh\beta
\left(\sqrt{\frac{\gamma}{2
x_0}} \sigma\kappa\right) \right],
\\
f(\sigma) &=& \frac{ u_0 M_P^2}{8 H_0^2} \,\e^{-\frac{u_0+1}{u_0}
\sqrt{\frac{\gamma}{2 x_0}}\sigma\kappa } {\rm sech}^2\beta
\left(\sqrt{\frac{\gamma}{2 x_0}} \sigma\kappa\right).
\end{eqnarray}
Here a natural choice of $u_0$ is $-1< u_0 <0$, so that the Hubble
parameter $H(\sigma)$ is a monotonically decreasing function of
the logarithmic time $\ln(a)$, or $N$. The effective potential is
given by
\begin{eqnarray}
\kappa^2 \Lambda(\sigma) &=& H_0^2 M_P^2 \,\e^{\frac{u_0+1}{u_0}
\sqrt{\frac{\gamma}{2 x_0}}\sigma\kappa } \cosh^2\beta
\left(\sqrt{\frac{\gamma}{2 x_0}}
\sigma\kappa\right) \nonumber \\
&{}& \quad \times \left(3-x_0-3 u_0-9 u_0
\left[h_1+\beta\tanh\beta\left(\sqrt{\frac{\gamma}{2 x_0}}
\sigma\kappa\right)\right]\right).
\end{eqnarray}
In theories of quintessence, one Taylor expands the potential
$\Lambda(\sigma) $ about the minimum, so as to obtain a dynamical
scale for the mass of the quintessence field. This potential
therefore merits further study.

\subsection{$u(N)=u_0$ and $h(N)=h_0$}

Consider, again, the case where $u=u_0$, but now instead of
$x(N)\simeq $ const, we demand that $h(N)=\dot{H}/H^2\simeq $
const. The solutions for $x(N)$ and $y(N)$ are given by
\begin{eqnarray}
x=-h_0 (1+u_0-h_0 u_0), \quad y=3+h_0(1-5u_0-h_0 u_0).
\end{eqnarray}
One also finds the following relationships:
\begin{equation}\label{rel-sigma-N}
\sigma= \pm \frac{\sqrt{2}}{\kappa}\sqrt{\frac{x}{\gamma}} N +
const, \quad H=H_0 \, \e^{h_0 N}.
\end{equation}
The effective potential is then given by
\begin{equation}
\Lambda(\sigma) = V_0 \e^{2h_0 N} \left(3+h_0 -3 u_0 - 8 h_0 u_0 -
h_0^2 u_0\right) \sim  \Lambda_0\,\e^{-\sigma/\sigma_0}.
\end{equation}

\subsection{$x(N)=x_0$ and $y(N)=y_0$}

Finally, as one more special case, let us consider that both the
potential term $V(\sigma)$ and the kinetic term
$K(\sigma)=\frac{\gamma}{2} \dot{\sigma}^2$ scale with $H^2$;
presumably, with different proportionality constants. For
$y=\kappa^2 \frac{V(\sigma)}{H^2}= y_0$ and $x=\kappa^2
\frac{K}{H^2}=x_0$, the solution is given by
\begin{eqnarray}
h(N)=-\frac{3+5x_0-y_0}{x_0+y_0+3}\equiv h_0, \quad
u(\sigma(N))=\frac{(3-x_0-y_0) h_0}{6} -u_1\,\e^{2h_0 N},
\end{eqnarray}
where $u_1$ is an integration constant. The effective potential is
therefore
\begin{eqnarray}
\Lambda(\sigma)  &=& V_0\,\e^{2 h_0 N} \left[ 3 u_1 (1+h_0) \e^{2
h_0 N}+2\left(y_0+\frac{x_0(x_0 -2
y_0-3)}{5x_0-y_0+3}\right)\right]\nonumber \\
&\simeq & \Lambda_0\,\e^{-\sigma/\sigma_0}+
\Lambda_1\,\e^{-2\sigma/\sigma_0}.
\end{eqnarray}
From the above results, it is clear that at late times, $\sigma\gg
\sigma_0$, the leading order contribution to the potential comes
from the term proportional to $\e^{-\sigma/\sigma_0}$.


\end{document}